%% file: ANA-EXOT-2019-03-PAPER.tex
\pdfoutput=1
\newcommand*{\ATLASLATEXPATH}{}
 
\newcommand{\AtlasCoordFootnote}{
ATLAS uses a right-handed coordinate system with its origin at the nominal interaction point (IP)
in the centre of the detector and the \(z\)-axis along the beam pipe.
The \(x\)-axis points from the IP to the centre of the LHC ring,
and the \(y\)-axis points upwards.
Cylindrical coordinates \((r,\phi)\) are used in the transverse plane,
\(\phi\) being the azimuthal angle around the \(z\)-axis.
The pseudorapidity is defined in terms of the polar angle \(\theta\) as \(\eta = -\ln \tan(\theta/2)\).
Angular distance is measured in units of \(\Delta R \equiv \sqrt{(\Delta\eta)^{2} + (\Delta\phi)^{2}}\).}
 
\documentclass[PAPER, atlasdraft=false, texlive=2016, UKenglish, subfigure, cernpreprint]{\ATLASLATEXPATH atlasdoc}
 
\usepackage{\ATLASLATEXPATH atlaspackage}
\usepackage{\ATLASLATEXPATH atlasbiblatex}
 
\usepackage{\ATLASLATEXPATH atlasphysics}
\usepackage{amssymb}
 
\graphicspath{{logos/}{figures/}}
 
\usepackage{ANA-EXOT-2019-03-PAPER-defs}

\AtlasTitle{Search for new resonances in mass distributions of jet pairs using 139~fb$^{-1}$~of $pp$ collisions at $\sqrt{s}=13$~TeV with the ATLAS detector}

\AtlasAbstract{
A search for new resonances decaying into a pair of jets is reported using the dataset of proton--proton collisions recorded at $\sqrt{s}=13$~TeV with the ATLAS detector at the Large Hadron Collider between 2015 and 2018, corresponding to an integrated luminosity of 139~fb$^{-1}$.
The distribution of the invariant mass of the two leading jets is examined for local excesses above a data-derived estimate of the Standard Model background.
In addition to an inclusive dijet search, events with jets identified as containing $b$-hadrons are examined specifically.
No significant excess of events above the smoothly falling background spectra is observed.
The results are used to set cross-section upper limits at 95\% confidence level on a range of new physics scenarios.
Model-independent limits on Gaussian-shaped signals are also reported.
The analysis looking at jets containing $b$-hadrons benefits from improvements in the jet flavour identification at high transverse momentum, which increases its sensitivity relative to the previous analysis beyond that expected from the higher integrated luminosity.
}
 
\author{The ATLAS Collaboration}
 
\AtlasRefCode{EXOT-2019-03}
 
\AtlasJournalRef{\JHEP 03 (2020) 145}
\AtlasDOI{10.1007/JHEP03(2020)145}

\PreprintIdNumber{CERN-EP-2019-162}

\AtlasCoverSupportingNote{Supporting document}{https://cds.cern.ch/record/2646455}
 
\AtlasCoverCommentsDeadline{3rd August 2019}
 
\AtlasCoverAnalysisTeam{
Yiming Abulaiti,
Benjamin William Allen,
Matteo Bauce,
Harinder Singh Bawa,
Iain Alexander Bertram,
Xin Chen,
Andrea Coccaro,
Francesco Armando Di Bello,
Wei Ding, Binbin Dong,
Simone Francescato,
Yongsheng Gao,
Stefano Giagu,
Douglas Gingrich,
Haifeng Li,
Bingxuan Liu,
Jeremy Love,
Nishu Nishu,
Trine Poulsen,
James Proudfoot,
Julian Schmoeckel,
David Strom,
Rui Wang,
Ben William Whitmore,
Philipp Windischhofer,
Dengfeng Zhang,
Ning Zhou}
 
\AtlasCoverEdBoardMember{James~Frost~(chair), Arely~Cortes-Gonzalez, Loic~Valery, Lailin~Xu}

\AtlasCoverEgroupEditors{atlas-EXOT-2019-03-editors@cern.ch}
 
\AtlasCoverEgroupEdBoard{atlas-EXOT-2019-03-editorial-board@cern.ch}

\hypersetup{pdftitle={ATLAS document},pdfauthor={The ATLAS Collaboration}}
 
\begin{document}
 
\maketitle
 
\section{Introduction}
Many models of physics beyond the Standard Model (SM) predict the existence of new heavy particles which couple to quarks and/or gluons.
Such heavy particles could be produced in proton\text{--}proton collisions at the Large Hadron Collider (LHC) and then decay into quarks and gluons, creating two energetic jets in the detector. In the SM, dijet events are produced mainly by quantum chromodynamics (QCD) processes. QCD predicts dijet events with a smoothly decreasing invariant mass distribution, \mjj.
A new particle decaying into quarks or gluons would emerge instead as a resonance in the \mjj spectrum.
 
If the new particle has a sizeable coupling to $b$-quarks and decays into $b\bar{b}$, $bq$ or $bg$ pairs, the identification of jets containing $b$-hadrons ($b$-tagging) in the decay final state could significantly enhance the sensitivity to such a new particle.
This analysis searches for resonant excesses in the \mjj distribution of the two most energetic jets with an inclusive jet selection and with separate selections where at least one or exactly two jets are identified as containing a $b$-hadron.
 
Dijet resonance searches have been performed at previous hadron colliders covering the dijet invariant mass range from $110~\GeV$ to $1.4~\TeV$~\cite{UA1_dijet,UA2_dijet,cdf_dijet,d0_dijet}.
At the LHC, the most recent searches probe masses up to $7.5~\TeV$~\cite{EXOT-2016-21,CMS-EXO-16-056}.
The lowest inspected \mjj value in the recent LHC searches is above 1~\TeV\ and is dictated by the trigger and data-acquisition systems of the experiments. Searching for resonances below the \TeV\ mass range is well motivated and alternative approaches employing more sophisticated trigger or analysis strategies have resulted in novel searches~\cite{EXOT-2016-20,EXOT-2018-05,EXOT-2017-01,CMS-EXO-16-030,CMS-EXO-17-001,CMS-EXO-17-024}. For new resonances decaying into jets containing $b$-hadrons, dedicated searches have been performed~\cite{EXOT-2016-33, CMS-EXO-16-057}.
 
In this analysis, the dataset recorded at $\sqrt{s}=13~\TeV$ with the ATLAS detector is used, corresponding to an integrated luminosity of 139~\ifb. The \mjj spectrum ranging from $1.1~\TeV$ to $8~\TeV$ is probed, and the results are interpreted in the context of several new physics scenarios, which include excited quarks $q^*$ ($q=(u,d,c,s,b)$) from compositeness models~\cite{qstar1,qstar2}; heavy \Zprime and \Wprime gauge bosons~\cite{zprime1,zprime3,wprime1}; a chiral excitation of the $W$ boson, denoted \Wstar~\cite{Chizhov:2009fc,Chizhov:2010jg}; a leptophobic \Zprime dark-matter mediator model~\cite{DM1,DM2,DM3}; quantum black holes ~\cite{qbh1,qbh2}; and Kaluza--Klein gravitons~\cite{RS1,RS2}.
In addition, limits on generic Gaussian-shaped narrow-resonance signals~\cite{EXOT-2013-11} are derived.
 
\section{ATLAS detector}
The ATLAS detector~\cite{PERF-2007-01} at the LHC covers nearly the entire solid angle around the collision point.\footnote{\AtlasCoordFootnote}
It consists of an inner tracking detector surrounded by a thin superconducting solenoid, electromagnetic and hadronic calorimeters, and a muon spectrometer incorporating three large superconducting toroidal magnets.
The inner-detector system is immersed in a \SI{2}{\tesla} axial magnetic field
and provides charged-particle tracking in the range \(|\eta| < 2.5\).
 
The high-granularity silicon pixel detector covers the vertex region and typically provides four measurements per track, the first hit normally being in the insertable B-layer installed before Run~2~\cite{ATLAS-TDR-2010-19,PIX-2018-001}.
It is followed by the silicon microstrip tracker which usually provides eight measurements per track.
These silicon detectors are complemented by the transition radiation tracker,
which enables radially extended track reconstruction up to \(|\eta| = 2.0\) and contributes to electron identification.
 
The calorimeter system covers the pseudorapidity range \(|\eta| < 4.9\).
Within the region \(|\eta|< 3.2\), electromagnetic calorimetry is provided by barrel and
endcap high-granularity lead/liquid-argon (LAr) calorimeters,
with an additional thin LAr presampler covering \(|\eta| < 1.8\),
to correct for energy loss in material upstream of the calorimeters.
Hadronic calorimetry is provided by the steel/scintillator-tile calorimeter,
segmented into three barrel structures within \(|\eta| < 1.7\), and two copper/LAr hadronic endcap calorimeters.
The solid angle coverage is completed with forward copper/LAr and tungsten/LAr calorimeter modules
optimised for electromagnetic and hadronic measurements, respectively.
 
The outermost layers of ATLAS consist of an external muon spectrometer within \(|\eta|< 2.7\),
incorporating three large toroidal magnet assemblies with eight coils each.
 
Interesting events were selected to be recorded by the first-level trigger system implemented in custom hardware, followed by selections made by algorithms implemented in software in the high-level trigger computer farm~\cite{TRIG-2016-01}.
The first-level trigger reduces the selection rate from the \SI{40}{\MHz} bunch crossing rate to below \SI{100}{\kHz},
which the high-level trigger further reduces in order to record events to disk at a rate of about \SI{1}{\kHz}.
 
\section{Simulated event samples}
Monte Carlo (MC) simulations are used to model the expected benchmark signals and to validate the SM background estimation.
 
In most of the sample generation, the leading-order (LO) NNPDF2.3 parton distribution functions (PDF)~\cite{Ball:2012cx} and the A14 Pythia tuned parameter set for the modelling of parton showers, hadronisation and the underlying event~\cite{ATL-PHYS-PUB-2014-021} were adopted, unless otherwise described below.
 
MC events from QCD multijet processes were generated with $\PYTHIA$~v8.186~\cite{Sjostrand:2007gs}.
The renormalisation and factorisation scales were set to the average transverse momentum \pt of the two leading (highest \pt) jets.
Generated events were reweighted to next-to-leading-order (NLO) predictions using \mjj-dependent correction factors~\cite{Nagy:2001fj,Nagy:2003tz,Catani:1996vz}.
To validate the modelling of the background, the MC simulation is normalized to the data and the shapes of various kinematic variables in simulation are compared with the data.
The MC simulation is found to agree with the data, with a difference of up to approximately 20\% in the tail regions.
 
Due to the limited size of the simulated samples and the large theoretical uncertainties of QCD processes, the background is estimated by fitting each of the data \mjj spectra as described in Section~\ref{sec:analysis}.
 
Several models of new physics were simulated, including heavy gauge bosons, a chiral excitation of the $W$ boson, excited quarks, quantum black holes and Kaluza--Klein gravitons.
The sequential standard model (SSM) \Zprime boson~\cite{zprime1}  has the same couplings to the SM fermions as the SM $Z$ boson, so the bottom-quark decay branching fraction $\mathcal{B}(\Zprime\to b\bar{b})$ is 13.8\%.
The intrinsic width of the SSM \Zprime boson is approximately $3\%$ of the resonance mass.
Events from the SSM \Zprime model were generated in the $b\bar{b}$ decay channel with $\PYTHIA$~v8.186 at LO, and the cross-sections were then corrected to the NLO predictions~\cite{Alwall:2014hca}.
 
A leptophobic \Zprime model with axial-vector couplings to SM quarks and containing a Dirac fermion dark matter (DM) candidate is considered~\cite{DM3}.
The events from \Zprime decaying into $q\bar{q}$ where $q=(u,d,s,c,b)$ were generated with \MGMCatNLO~2.4.3~\cite{Alwall:2014hca} with the DM mass fixed to $10~\TeV$ and the coupling to dark matter ($g_\chi$) set to 1.5.
The mediator \Zprime mass ranges from $1~\TeV$ to $7~\TeV$, and the coupling to SM quarks ($g_\text{q}$) varies from 0.1 to 0.5.
In this scenario, the \Zprime does not decay into the DM candidate and so the dijet signal depends only upon the coupling to quarks and the mass of the \Zprime resonance.
The chosen $g_\text{q}=0.5$ coupling corresponds to a width of 12\% of the resonance mass, nearly the maximum width to which this search is sensitive.
For the resonance searches with $b$-tagging, dedicated samples of $\Zprime$ signals decaying into $b\bar{b}$ final states were simulated using the same generator set-up as for the inclusive samples. In this leptophobic case, the bottom-quark decay branching fraction $\mathcal{B}(\Zprime\to b\bar{b})$ is 18.9\%.
 
A heavy charged \Wprime gauge boson model~\cite{wprime1} with $V-A$ couplings was simulated similarly to the SSM \Zprime scenario, using $\PYTHIA$~v8.186 at LO. The mass of the \Wprime ranges from $1~\TeV$ to $6.5~\TeV$ and only hadronic decays of the \Wprime were simulated, with all six quark flavours included.
 
Events with a chiral excitation of the $W$ boson, \Wstar, arising from a $W$ compositeness model~\cite{Chizhov:2009fc,Chizhov:2010jg}, were generated with \textsc{CalcHep}~v3.6~\cite{CalcHEP}, and then processed with $\PYTHIA$~v8.210 for the simulation of non-perturbative effects.
The angular distribution of the decay products differs strongly from those of all other models considered in this analysis and has an excess more towards the forward region, which motivates using a different kinematic selection for this signal. The decays of the $W^*$ were set to be leptophobic and include all SM quarks.
Event samples of $W^*$ bosons were generated with masses ranging from $1.8~\TeV$ to $6.0~\TeV$.
 
Excited quark ($q^*$) signal samples~\cite{qstar1,qstar2} were generated with $\PYTHIA$~v8.186, assuming spin-\textonehalf\ excited quarks with the same coupling constants as SM quarks. Both light flavour $(u,d,s)$ and heavy flavour $(c,b)$ quarks were taken into account in the event generation.
The generated $q^*$ masses range from $2~\TeV$ to $8~\TeV$. The compositeness scale was set to the excited-quark mass. Only the decay into a gluon and an up- or down-type quark was simulated; this is the dominant process in the dijet final state, with a branching ratio of 85\%.
Excited $b$-quark ($b^*$) signal samples were produced specifically for searches in the $b$-tagged dijet categories.
The same mass range as for the $q^*$ signal samples was simulated with analogous generator settings.
All decay modes were simulated with the dominant mode being the $bg$ channel, with a branching fraction of 85\%, and the remaining decay modes being $b\gamma$, $bZ$ and $tW$.
 
In models with large extra dimensions~\cite{ADD}, the fundamental scale of gravity $M_\text{D}$ is lowered to a few$~\TeV$.
Quantum black holes (QBH)~\cite{qbh1,qbh2}, the quantum analogues of ordinary black holes, can be produced at or above this scale at the LHC.
Once produced, QBH would decay into two-body final states, mainly jets.
Events from a QBH model were generated with \textsc{BlackMax}~\cite{Dai:BlackMax} for six extra dimensions, using the CTEQ6L1 PDF set~\cite{Pumplin:2002vw} and with $M_\text{D}$ ranging from $4~\TeV$ to $10~\TeV$.
 
In the Randall\text{--}Sundrum extra dimension model~\cite{RS1,RS2}, the Kaluza\text{--}Klein (KK) spin-2 graviton decays preferentially into gluons and quarks.
Graviton signal samples were generated with $\PYTHIA$~v8.212 assuming the curvature parameter $k/\overline{M}_\text{PL}=0.2$, where $\overline{M}_\text{PL}$ is the four-dimensional reduced Planck scale.
The KK graviton samples were simulated in the $G \rightarrow b\bar{b}$ decay mode, with masses ranging from $1.25~\TeV$ to $7~\TeV$.
 
The generated background samples from QCD processes were passed through a full ATLAS detector simulation~\cite{SOFT-2010-01} using \GEANT4~\cite{Agostinelli:2002hh}.
The signal MC samples were passed through a fast simulation which relies on a parameterisation of the calorimeter response~\cite{ATL-PHYS-PUB-2010-013}.
The decay of $b$- and $c$-hadrons was performed consistently using the \textsc{EvtGen v1.2.0} decay package~\cite{Lange:2001uf}.
To account for additional proton\text{--}proton interactions (pile-up) from the same and neighbouring bunch crossings, a number of inelastic $pp$ interactions were generated with $\PYTHIA$~v8.186 using the~NNPDF23LO~PDF set~\cite{Martin:2009iq} and the ATLAS A3 set of tuned parameters~\cite{ATL-PHYS-PUB-2016-017}. These events were then superimposed onto the hard-scattering events.
All simulated events were weighted so that the distributions of the average number of collisions per bunch crossing in simulation and in data match.
 
\section{Data and event selection}
The data for this analysis were collected by the ATLAS detector from $pp$ collisions at the LHC with a centre-of-mass energy of $\sqrt{s}=13~\TeV$ in the years from 2015 to 2018.
With requirements that all detector systems were functional and recording high-quality data, the dataset corresponds to an integrated luminosity of 139~\ifb.
The uncertainty in the combined 2015\text{--}2018 integrated luminosity is 1.7\% \cite{ATLAS-CONF-2019-021}, obtained using the LUCID-2 detector \cite{LUCID2} for the primary luminosity measurements.
Events are selected using a trigger that requires at least one jet with \pt greater than $420~\GeV$, the lowest-\pt non-prescaled single-jet trigger.
 
Collision vertices are reconstructed from at least two tracks with $\pt>0.5$~GeV. The primary vertex is selected as the one with the highest $\sum p^2_{\text{T}}$ of the associated tracks.
 
In event reconstruction, calorimeter cells with an energy deposit significantly above the calorimeter noise are grouped together according to their contiguity to form topological clusters~\cite{PERF-2014-07}. These are then grouped into jets using the \antikt algorithm~\cite{Cacciari:2008gp, Fastjet} with a radius parameter of $R$~=~0.4. Jet energies and directions are corrected by jet calibrations as described in Ref.~\cite{PERF-2016-04}. Events are rejected if any jet with $\pt>150~\GeV$ is compatible with noise bursts, beam-induced background or cosmic rays using the `loose' criteria defined in Ref.~\cite{ATLAS-CONF-2015-029}.
 
Jets containing a $b$-hadron are identified using a deep-learning neural network, DL1r, for the first time at ATLAS. The DL1r $b$-tagging is based on distinctive features of $b$-hadrons in terms of the impact parameters of tracks and the displaced vertices reconstructed in the inner detector.
The inputs of the DL1r network also include discriminating variables constructed by a recurrent neural network (RNNIP)~\cite{ATL-PHYS-PUB-2017-003}, which exploits the spatial and kinematic correlations between tracks originating from the same $b$-hadron.
This approach is
found chiefly to improve the performance for jets with high~\pt~\cite{ATL-PHYS-PUB-2017-013}.
Operating points are defined by a single cut-value on the discriminant output distribution and are chosen to provide a specific $b$-jet efficiency for an inclusive \ttbar MC sample.
A 77\% efficiency $b$-tagging operating point is adopted, which gives maximal overall signal sensitivity across the various signal models and masses considered in the $b$-tagged categories.
The $b$-tagging performance has a strong dependence on the jet~\pt: the efficiency drops from 65\% for a $b$-jet~\pt\ of around $500~\GeV$ to 10\% for~a~\pt\ of around $2~\TeV$. Estimated from MC simulation, the corresponding mis-tag rate of charm jets drops from 15\% to 2\% over the same~\pt\ interval, and that of light-flavour jets remains at the level of 1\%.
Simulation-to-data scale factors are applied to the simulated event samples to compensate for differences in the $b$-tagging efficiency between data and simulation.
These scale factors are measured as a function of jet~\pt\ using a likelihood-based method in a sample highly enriched in \ttbar events~\cite{FTAG-2018-01}.
Given that the number of $b$-jets in data is limited for jet~\pt\ $>400~\GeV$, additional uncertainties are assessed by varying in the simulation the underlying quantities that are known to affect the $b$-tagging performance.
The differences between the $b$-tagging efficiency after each variation and the nominal $b$-tagging efficiency are then used to construct an extrapolation uncertainty to extend the validity of the correction factors into the higher jet-\pt\ range used in this analysis.
The simulation-to-data scale factor as a function of jet \pt for the 77\% operating point of the DL1r $b$-tagging algorithm adopted in this search is shown in Figure~\ref{fig:SF}. More details about the procedure for the extraction and extrapolation of the $b$-tagging scale factors can be found in Ref.~\cite{FTAG-2018-01}.

\begin{figure}[thbp]
\centering
\includegraphics[width=0.6\textwidth]{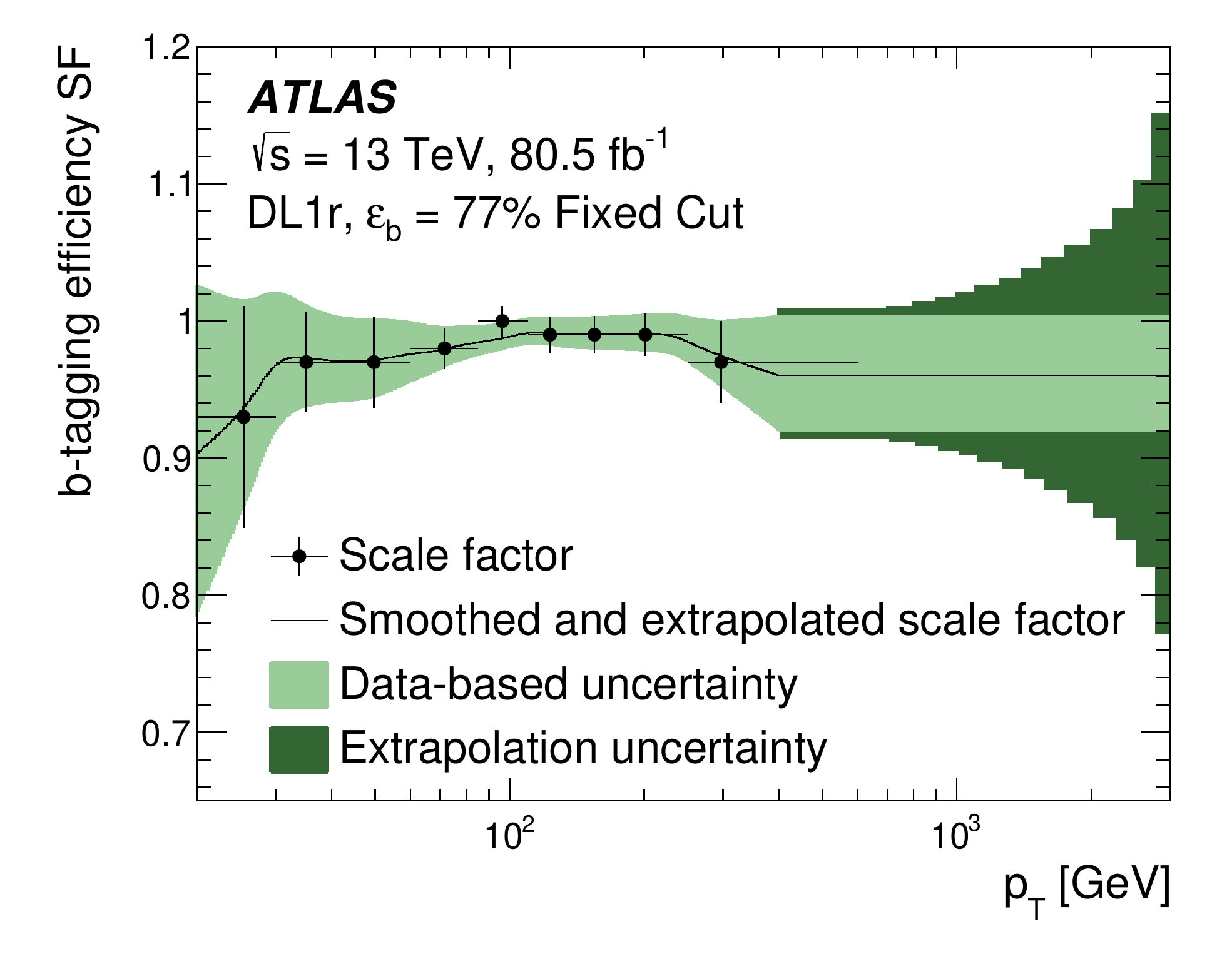}
\caption{
Simulation-to-data scale factor as a function of jet \pt for the 77\% operating point of the DL1r $b$-tagging algorithm.
The scale factors are measured with a likelihood-based method in a sample highly enriched in \ttbar events using 2015\text{--}2017 data, as described in Ref.~\cite{FTAG-2018-01}, with uncertainties due to the limited size of data sample, detector calibration and physics modelling.
An additional uncertainty is included to extrapolate the measured uncertainties to the high-\pt region of interest ($\pt > 400~\GeV$), and has contributions related to the reconstruction of tracks and jets, the modelling of the $b$-hadrons and the interaction of long-lived $b$-hadrons with the detector material.
}
\label{fig:SF}
\end{figure}
 
The analysis selections and the corresponding signal models investigated are summarised in Table~\ref{tab:selection}.
Events must contain at least two jets with \pt greater than $150~\GeV$ and the azimuthal angle between the two leading jets must be greater than 1.0.
To maximise the sensitivities to various signal models, the events are classified into an inclusive category with no $b$-jet tagging requirement, a one-$b$-tagged category ($1b$), requiring at least one of the two leading jets to be $b$-tagged, and a two-$b$-tagged category ($2b$), with both of the two leading jets being $b$-tagged.
For categories selecting $b$-jets, the two leading jets must be within $|\eta|<2.0$.
 
To reduce the dominant background contribution from QCD processes, a selection based on half of the rapidity separation between the two leading
jets, $y^*=(y_1-y_2)/2$, is implemented, where $y_1$ and $y_2$ are the rapidities of the leading jet and subleading jet respectively. The signal dijet events are produced through $s$-channel processes, which favour small $|y^*|$, while a large fraction of the background events are from QCD $t$-channel processes and have large $|y^*|$. The $|y^*|$ cut values are optimised for various categories and signals. In the inclusive selection, $|y^*|<0.6$ is required for the considered signals, except $W^*$. Due to the fact that a larger $|y^*|$ is favoured in the $W^*$ decays, a looser requirement $|y^*|<1.2$ is adopted in the search for $W^*$ signals. In the $b$-tagged categories, where the two leading jets have $|\eta|<2.0$, a selection $|y^*|<0.8$ is made.
 
A lower bound on the dijet invariant mass \mjj is required to ensure a fully efficient selection without any kinematic bias; it is determined by the single-jet trigger's efficiency turn-on and also depends on the $|y^*|$ requirement, as shown in Table~\ref{tab:selection}.
Within the acceptance of the \mjj and $|y^*|$ selections, the leading jet's \pt is above the single-jet trigger's threshold.
For the inclusive selection, the acceptance of QBH and $q^*$ signals is around 55\% for all the masses considered, while that of $W'$ and $\Zprime$ ranges from approximately 20\% to 45\%, depending on the resonance mass. For the $W^*$ selection, the acceptance increases from 30\% to 70\% for $W^*$ mass values from $2~\TeV$ to $6~\TeV$. For the $b$-tagged categories, the acceptance of $b^*$ and $\Zprime(b\bar{b})$ increases from 20\% and reaches a plateau of around 70\% at a mass of 2.5~\TeV.
 
The signal selection efficiencies from the $b$-tagging requirement (per-event $b$-tagging efficiencies) shown in Figure~\ref{fig:beff} are derived after applying the rest of the event selection. The efficiency decreases as \mjj increases, since the b-tagging efficiency decreases when the jet \pt increases.
In the $1b$ category, the efficiency for final states containing two $b$-quarks, such as a \Zprime signal, is higher than for the $b^*$ signal. At high mass, because the gluon from the $b^*$ decay is more likely to split into a $b\bar{b}$ pair, the per-event $b$-tagging efficiency of the $b^*$ signal is enhanced and closer to what is observed in simulated \Zprime events.
 
\begin{figure}[htbp]
\centering
\includegraphics[width=0.65\textwidth]{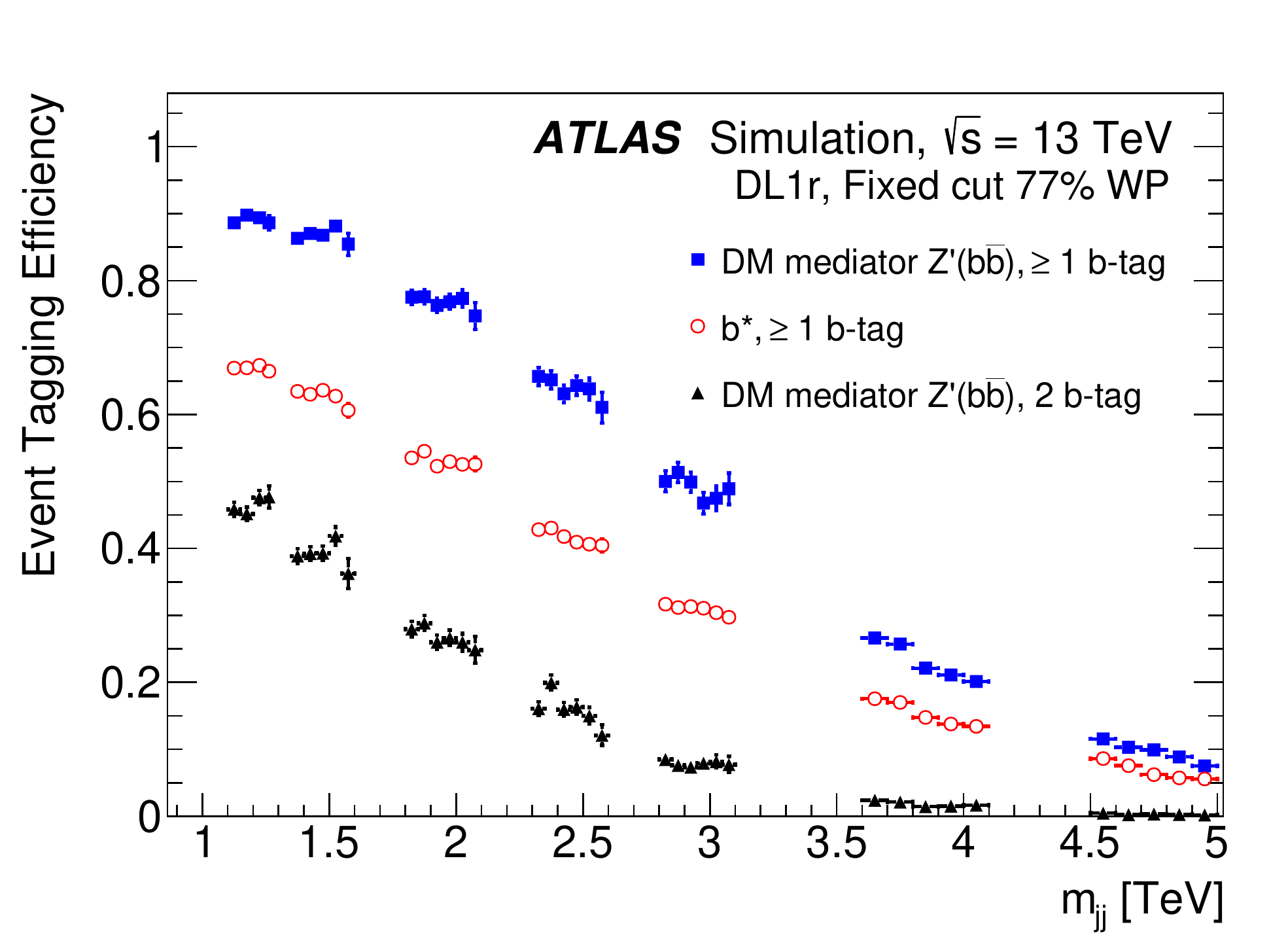}
\caption{The probability of an event to pass the $b$-tagging requirement after the rest of the event selection, shown as a function of the resonance mass \mjj and for the $1b$ and $2b$ analysis categories.}
\label{fig:beff}
\end{figure}

\begin{table}[htbp]
\caption{Summary of the event selection requirements and benchmark signals being tested in each analysis category. Only the two jets with highest \pt enter in the event selection. The exact values of the \mjj lower bounds also depend on the jet energy resolution uncertainty.}
\centering
\begin{tabular}{l|c|c|c|c}
\hline\hline
Category & \multicolumn{2}{c|}{Inclusive} & $1b$ & $2b$ \\\hline
Jet \pt & \multicolumn{4}{c}{$>150~\GeV$} \\\hline
Jet $\phi$ & \multicolumn{4}{c}{$|\Delta \phi(jj)|>1.0$} \\\hline
Jet $|\eta|$& \multicolumn{2}{c|}{-} &  \multicolumn{2}{c}{$<2.0 $} \\\hline
$|y^*|$ & $<0.6$ &  $<1.2$ &  \multicolumn{2}{c}{$<0.8$} \\\hline
$\mjj$ & $>1100~\GeV$ & $>1717~\GeV$ & \multicolumn{2}{c}{$>1133~\GeV$} \\\hline
$b$-tagging & \multicolumn{2}{c|}{no requirement} & $\geqslant1$ $b$-tagged jet & 2 $b$-tagged jets \\\hline
\multirow{5}{*}{Signal} & DM mediator \Zprime & $W^*$ & $b^*$ & DM mediator \Zprime ($b\bar b$) \\
&\Wprime & & Generic Gaussian & SSM \Zprime ($b\bar b$) \\
& $q^*$  & &                 & graviton ($b\bar{b}$) \\
& QBH    & &                 & Generic Gaussian \\
& Generic Gaussian & &  & \\\hline
\hline
\end{tabular}
\label{tab:selection}
\end{table}

\section{Dijet mass spectrum}
\label{sec:analysis}
The SM production of dijet events is dominated by QCD multijet processes, which yield a smoothly falling \mjj spectrum. To determine the SM contribution, the sliding-window fitting method~\cite{EXOT-2016-21} is applied to the data, with a nominal fit using a parametric function:
 
\begin{equation*}
f(x)=p_1(1-x)^{p_2}x^{p_3+p_4\ln x}
\end{equation*}
 
where $x= \mjj/\sqrt{s}$ and $p_{1,2,3,4}$ are the four fitting parameters.
The background in each \mjj bin is extracted from the data by fitting in a mass window centred around that bin. The window size is chosen to be the largest possible window that satisfies the fit requirements described later in this section.
 
Several data-driven background \mjj spectra are used to validate the background fitting strategy.
On these spectra, `signal injection tests' and `spurious signal tests' are performed to validate the sliding-window fit.
For the $b$-tagged categories, the background-only spectra are derived from control regions (CRs) which are constructed by reversing the requirement on $|y^*|$ or removing the $b$-tagging requirement.
In these CRs the signal leakage is expected to be small, and this is confirmed by the MC simulation.
In the CRs with the $|y^*|<0.8$ requirement reversed, per-event fractions passing $b$-tagging selections are derived as functions of \pt and $\eta$ of the two leading jets for both the $1b$ and  $2b$ categories, which fully take into account the correlations between the leading and subleading jets.
The dijet spectra from QCD processes in the $b$-tagged signal regions are obtained from the CR with no $b$-tagging requirement (using the signal region $|y^*|$ selection), multiplied by the appropriate $b$-tagging efficiencies.
For the inclusive category, in the absence of a background-dominated control region, a test spectrum corresponding to an integrated luminosity of 139~\ifb\ is created to perform these tests by scaling up the background-only fit to the 37~\ifb\ dataset, which is already published in Ref.~\cite{EXOT-2016-21} with no evidence of new physics, and then fluctuating the content of each bin around the fit value according to a Poisson distribution.
No significant bias is observed in the tests, as described below.
 
In the signal injection tests, various signal models are added to the expected background distribution to assess whether or not the sliding-window procedure is able to fit the combined distribution and measure the correct signal yield. This test is designed to evaluate how sensitive the sliding-window fit is to all the tested signal types. For each of the benchmark and Gaussian-shaped signals, the extracted signal yield is consistent with that injected within the statistical uncertainty.
 
In the spurious signal tests, signal-plus-background fits are run on the background-only spectra for different signal masses and the extracted signal yield is taken as an estimate of the spurious signal. This test evaluates the robustness of the background fitting strategy and the capability of the fit function to model the background.
All signals considered for the inclusive categories show no bias, with the exception of Gaussian-shaped resonances with relative widths of 15\% where a spurious signal yield of up to 12\% of the statistical uncertainty of the estimated background from the fit is observed at high mass, where data counts are limited. In the $b$-tagged categories, the spurious signal yield observed for all the signals considered is between 10\% and 20\% of the statistical uncertainty of the estimated background fit.
A corresponding systematic uncertainty is assigned for affected signals as described in Section~\ref{sec:sys}.
 
The statistical significance of any localised excess in the \mjj distribution is quantified using the \BumpHunter test~\cite{Aaltonen:2008vt,Choudalakis:2011bh}. The \BumpHunter calculates the significance of any excess found in continuous mass intervals in all possible locations of the binned \mjj distribution. The search window's width varies from a minimum of two \mjj mass bins up to half the extent of the full \mjj mass distribution.
For each interval in the scan, \BumpHunter computes the significance of the difference between the data and the background.
The interval that deviates most significantly from the smooth spectrum is defined by the set of bins that have the smallest probability of arising from a Poisson background fluctuation. The probability of random fluctuations in the background-only hypothesis to create an excess at least as significant as the one observed anywhere in the spectrum, the \BumpHunter $p$-value, is determined by performing a series of pseudo-experiments drawn from the background estimate, with the look-elsewhere effect~\cite{lookelsewhere} considered. The fitting quality is assessed via the \BumpHunter $p$-value. In a good fit, any localised excess is expected to arise from fluctuations in the fitted background distribution. In determining the window size of the sliding-window fit, a fit is accepted if the corresponding \BumpHunter $p$-value is greater than 0.01.
 
Figure~\ref{fig:dijetspectra} shows the observed \mjj distributions for the various categories.
The bin widths for each category are chosen to approximate the \mjj resolution, which broadens with increasing \mjj mass.
Predictions for benchmark signals are scaled to larger cross-sections, from 10 to 1000 times their expected values, for display purposes.
The vertical lines indicate the most discrepant interval identified by the \BumpHunter test.
No significant deviation from the background-only hypothesis is observed in the data spectra.
In the inclusive category, the \BumpHunter $p$-values of the most discrepant regions are 0.89 for dijet events with $|y^*|<0.6$ and 0.88 for events with $|y^*|<1.2$. In the $b$-tagged categories, the \BumpHunter $p$-values of the most discrepant regions are 0.69 for $1b$ and 0.83 for $2b$.
The lower panel in each plot of Figure~\ref{fig:dijetspectra} shows the significance of the bin-by-bin differences between the data and the fit, as calculated from Poisson probabilities, considering only statistical uncertainties.

\begin{figure}[htbp]
\centering
\subfigure[][]{
\includegraphics[width=0.49\textwidth]{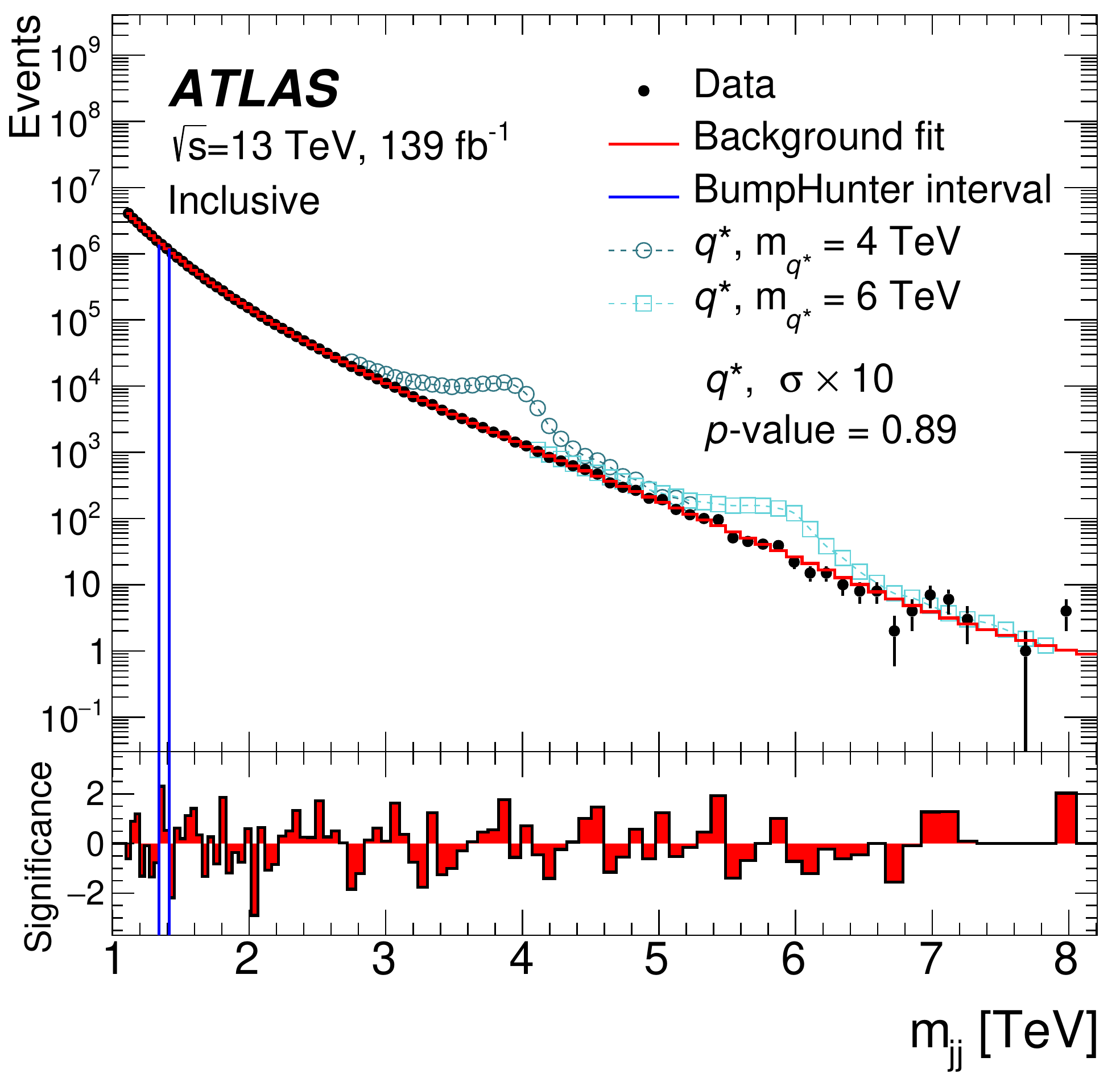}
}\subfigure[][]{
\includegraphics[width=0.49\textwidth]{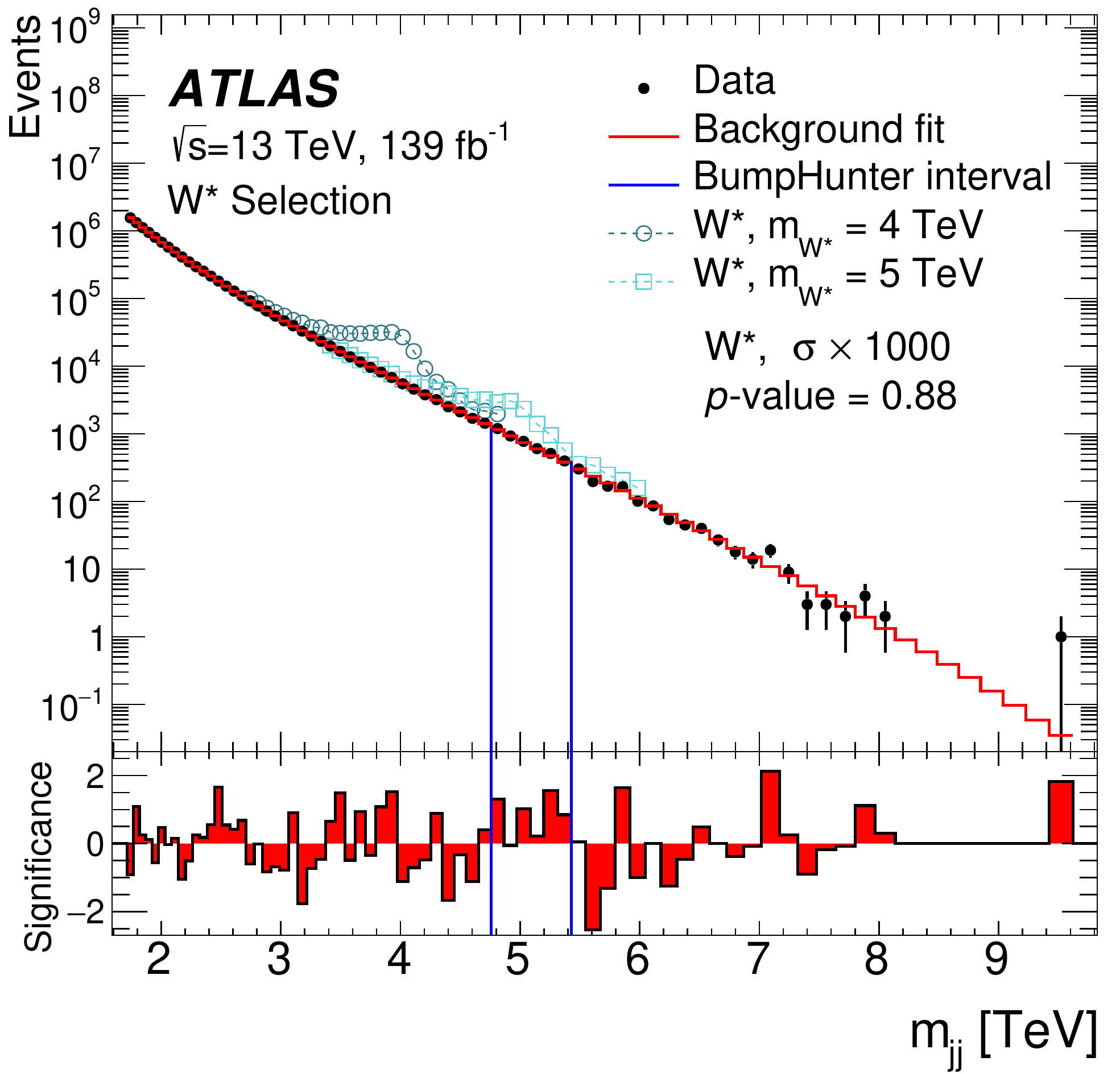}
} \\
\subfigure[][]{
\includegraphics[width=0.49\textwidth]{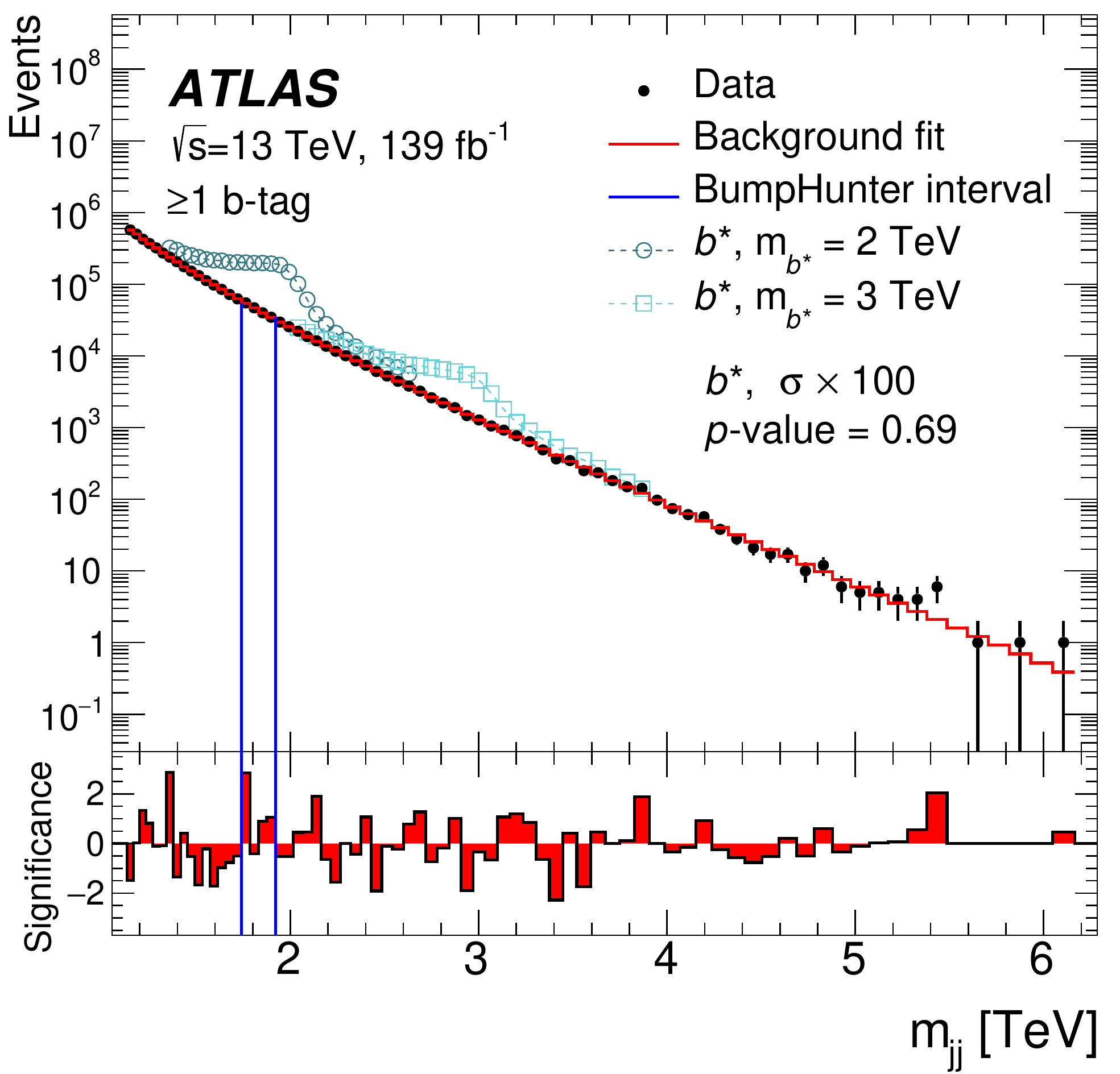}
}\subfigure[][]{
\includegraphics[width=0.49\textwidth]{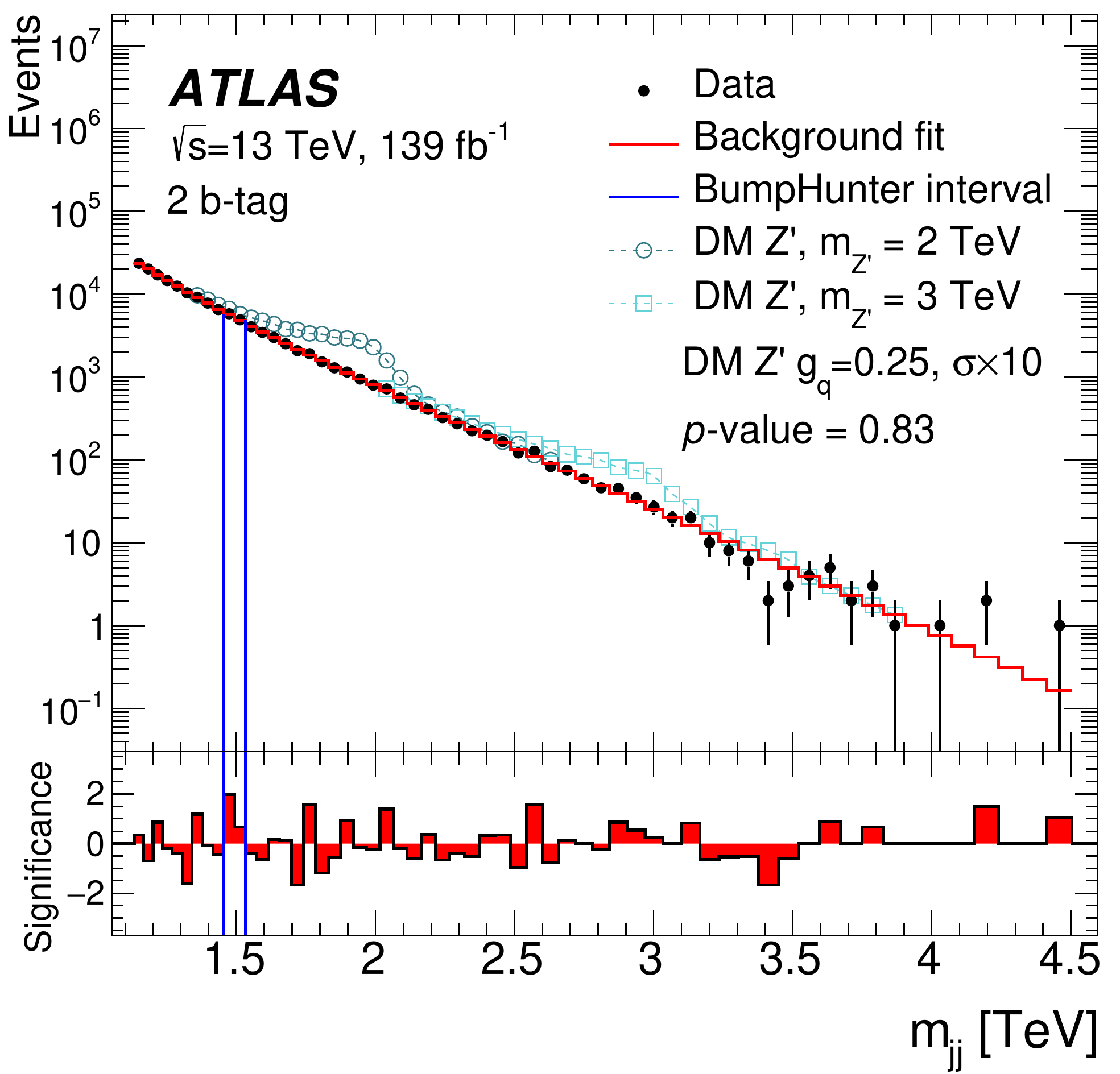}
}
\caption{
Dijet invariant mass distributions from multiple categories: (a) inclusive dijet with $|y^*|<0.6$, (b) inclusive dijet with $|y^*|<1.2$, (c) dijet with at least one $b$-tagged jet and (d) dijet with both jets $b$-tagged.
The vertical lines indicate the most discrepant interval identified by the \BumpHunter test, for which the $p$-value is stated in the figure.
}
\label{fig:dijetspectra}

\end{figure}
 
\section{Systematic uncertainties}
\label{sec:sys}
The statistical uncertainty of the fit due to the limited size of the data sample and the uncertainty due to the choice of fit function are considered as systematic uncertainties affecting the data-driven background determination.
 
To estimate these uncertainties, a large number of pseudo-data sets ($\sim 10\,000$) are generated as Poisson fluctuations from the nominal distribution. The statistical uncertainty in the values of the parameters in the fit function is derived by repeating the sliding-window fitting procedure on the pseudo-data. The uncertainty in each \mjj bin is taken to be the root mean square of the fit results in that bin for all pseudo-experiments, which increases from approximately 0.1\% at $\mjj=2~\TeV$ to 30\%\text{--}40\% in the high \mjj tail region. These uncertainties, and the ones throughout this Section, are expressed as variations relative to the nominal values.
 
The uncertainty due to the choice of background parameterisation is estimated by fitting the pseudo-data with the nominal function and alternative parametric functions. To determine the alternative functional form, several fits are performed using variations of the nominal function with at most one additional free parameter. The functional form used to estimate the systematic uncertainty is taken as the function giving the largest difference from the nominal fit while still fulfilling the fit quality criteria.
For the inclusive category, the alternative function has the form $p_1(1-x)^{p_2}x^{p_3+p_4\ln x+p_5x}$ while for the $b$-tagged categories, where the $b$-tagging efficiency biases the \mjj distribution, the form $p_1(1-x)^{p_2+p_3 x}x^{p_4 +p_5\ln x}$ is adopted.
The difference between the alternative background prediction and the nominal one, averaged across the set of pseudo-data, is considered as a systematic uncertainty, which reaches 10\% in the highest mass regions investigated in this analysis.
 
An additional systematic uncertainty is considered, based on the spurious signal tests.
In the inclusive category, this systematic uncertainty is required only for the Gaussian-shaped signal with a width of 15\% of its mass, since for the other signal hypotheses no bias is seen.
For the $b$-tagged categories, this uncertainty is considered for each signal according to the size of the observed effect.
The effect of this uncertainty on the signal cross-sections is found to be less than 5\% of the excluded values for all benchmark and Gaussian-shaped signals considered.
 
The main systematic uncertainties in the MC signal samples include those associated with the modelling of the jet energy scale (JES), the jet energy resolution (JER) and the $b$-tagging efficiency. JES and JER variations are applied to all the signals and affect the signal templates.
They are estimated using jets in $13~\TeV$ data and simulation in various methods as described in Ref.~\cite{PERF-2016-04}.
The JES uncertainty is less than 2\% of the jet~\pt for dijet invariant mass below $5~\TeV$ and around 4\% for higher mass.
The JER uncertainty ranges from 3\% to 6\% across the whole dijet invariant mass range investigated.
 
In the categories selecting one or two jets from $b$-hadrons, the systematic uncertainty of the $b$-tagging efficiency dominates. The uncertainty is measured using data enriched in $t\bar{t}$ events for jet $\pt<400~\GeV$ and extrapolated to higher-\pt regions~\cite{FTAG-2018-01}.
Dedicated simulations are used to extrapolate the measured uncertainties to the high-\pt\ region of interest.
Contributions related to the reconstruction of tracks and jets, the modelling of the $b$-hadrons and the interaction of long-lived $b$-hadrons with the detector material are considered.
Among the uncertainties associated with the reconstruction of tracks, those found to affect the $b$-tagging performance the most are the ones related to the track impact-parameter resolution, the fraction of fake tracks, the description of the detector material, and the track multiplicity per jet.
The uncertainty increases from 2\% for a jet~\pt of around $90~\GeV$ to 20\% for a jet \pt of around $3~\TeV$.
The overall $b$-tagging uncertainty affecting the normalisation of the Gaussian-shaped signals is taken into account.
 
A luminosity uncertainty of 1.7\% is applied to the normalisation of the signal samples.
Uncertainties in the signal acceptance associated with the choice of PDF and the scale choices are found to be approximately 1\% for most signals, reaching 4\% for high mass values.
 
\section{Signal interpretation}
Since no significant deviation from the expected background is observed, constraints on various signal models that would produce a resonance in the dijet invariant mass distribution are derived using a frequentist framework~\cite{Baak:2014wma}.
Upper limits on the signal cross-section times acceptance times branching ratio are extracted at 95\% confidence level (CL) using the CL$_\text{s}$ method~\cite{Read:2002hq} with a binned profile likelihood ratio as the test statistic.
For the $1b$ and $2b$ categories, the upper limits are set on the signal cross-section times acceptance times $b$-tagging selection efficiency times branching ratio.
The expected limits are calculated with the asymptotic approximation to the test statistic's distribution~\cite{Cowan:2010js} and using pseudo-experiments generated according to the values of the background uncertainties from the maximum-likelihood fit.
Pseudo-experiments are employed for the interpretation of the signals populating the high-mass part of the spectra where the relative deviation from the asymptotic approximation is found to be more than 1\%.
The calculated limits are logarithmically interpolated.
No uncertainty is applied to the signal theoretical cross-sections.
The systematic uncertainties of the background and signal samples are incorporated into the limits by varying all the uncertainty sources according to Gaussian probability distributions.
For the signal models considered here, the new physics resonance's couplings are strong compared with the scale of perturbative QCD at the signal mass, so that the interference with QCD terms can be neglected.
 
\begin{figure}[!thbp]
\centering
\subfigure[][]{
\includegraphics[width=0.49\textwidth]{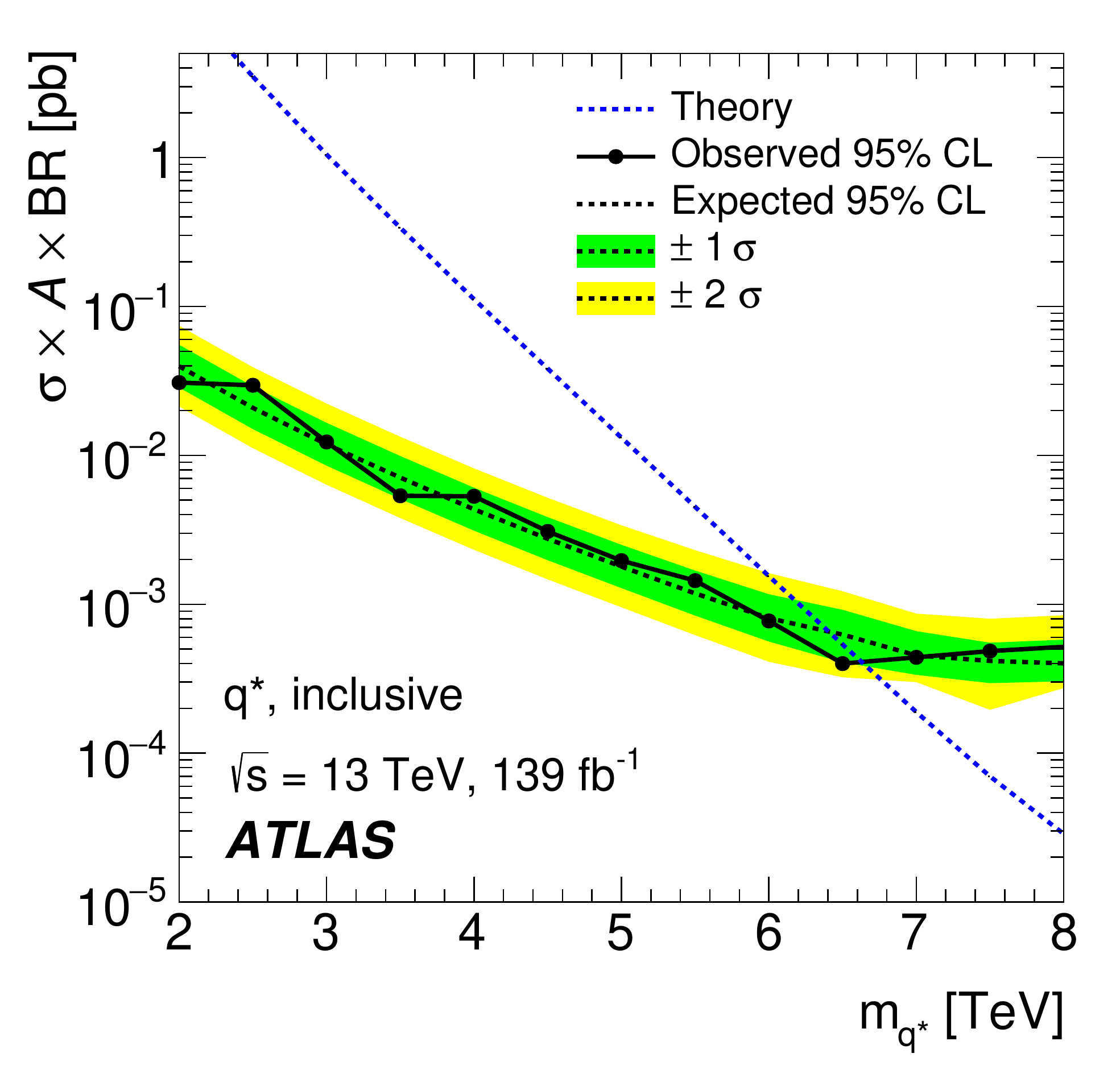}
}\subfigure[][]{
\includegraphics[width=0.49\textwidth]{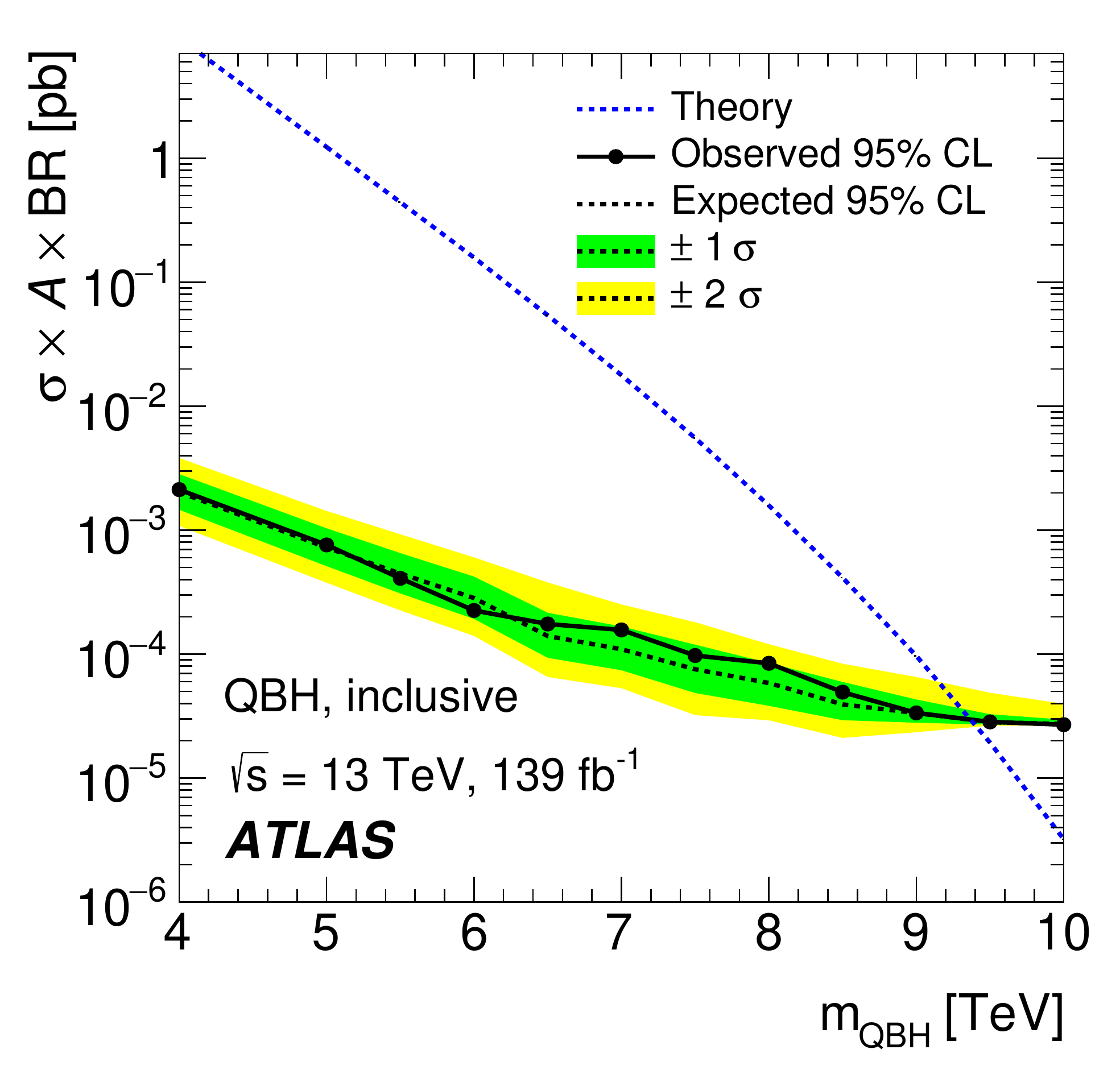}
} \\
\subfigure[][]{
\includegraphics[width=0.49\textwidth]{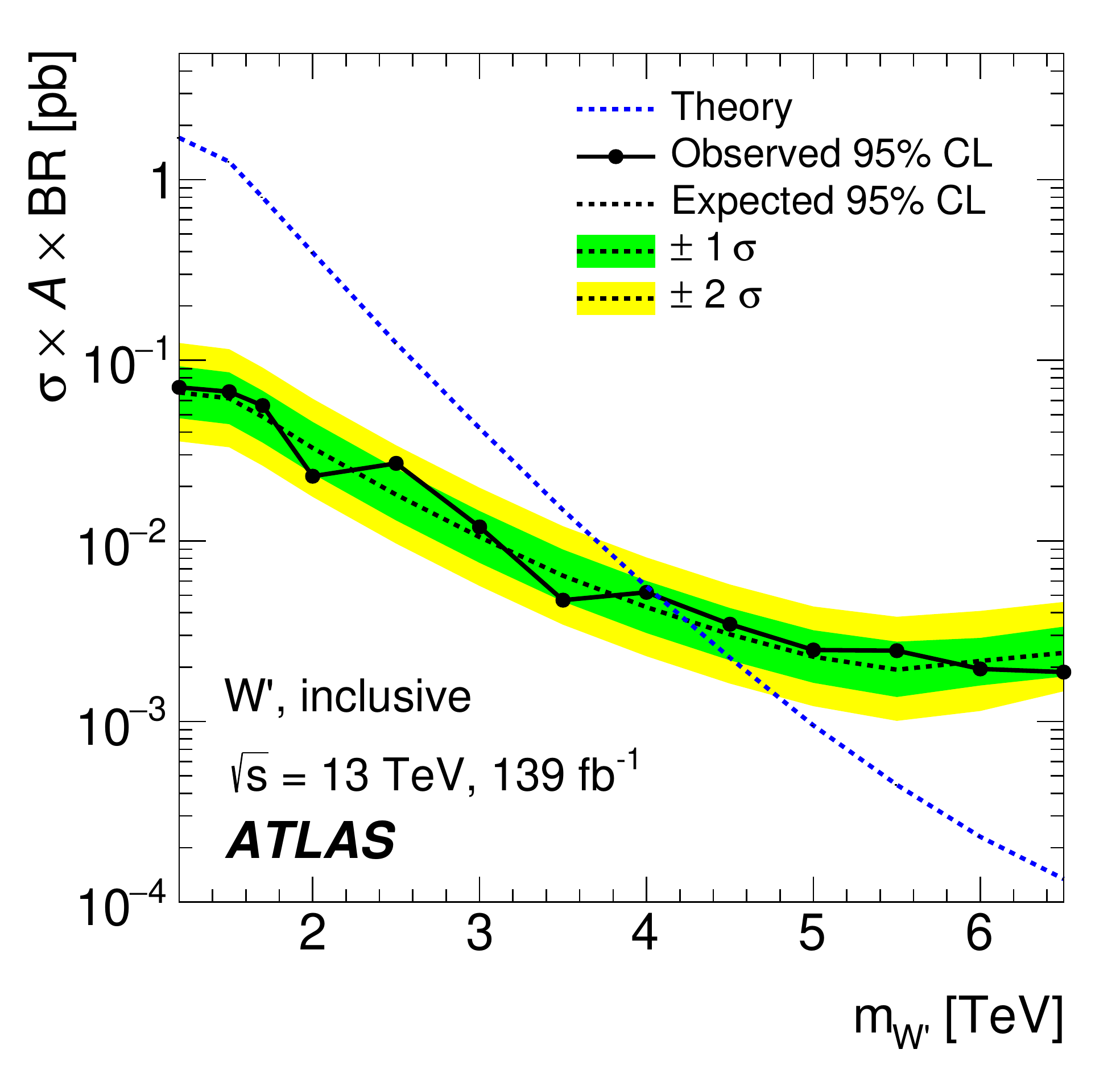}
}\subfigure[][]{
\includegraphics[width=0.49\textwidth]{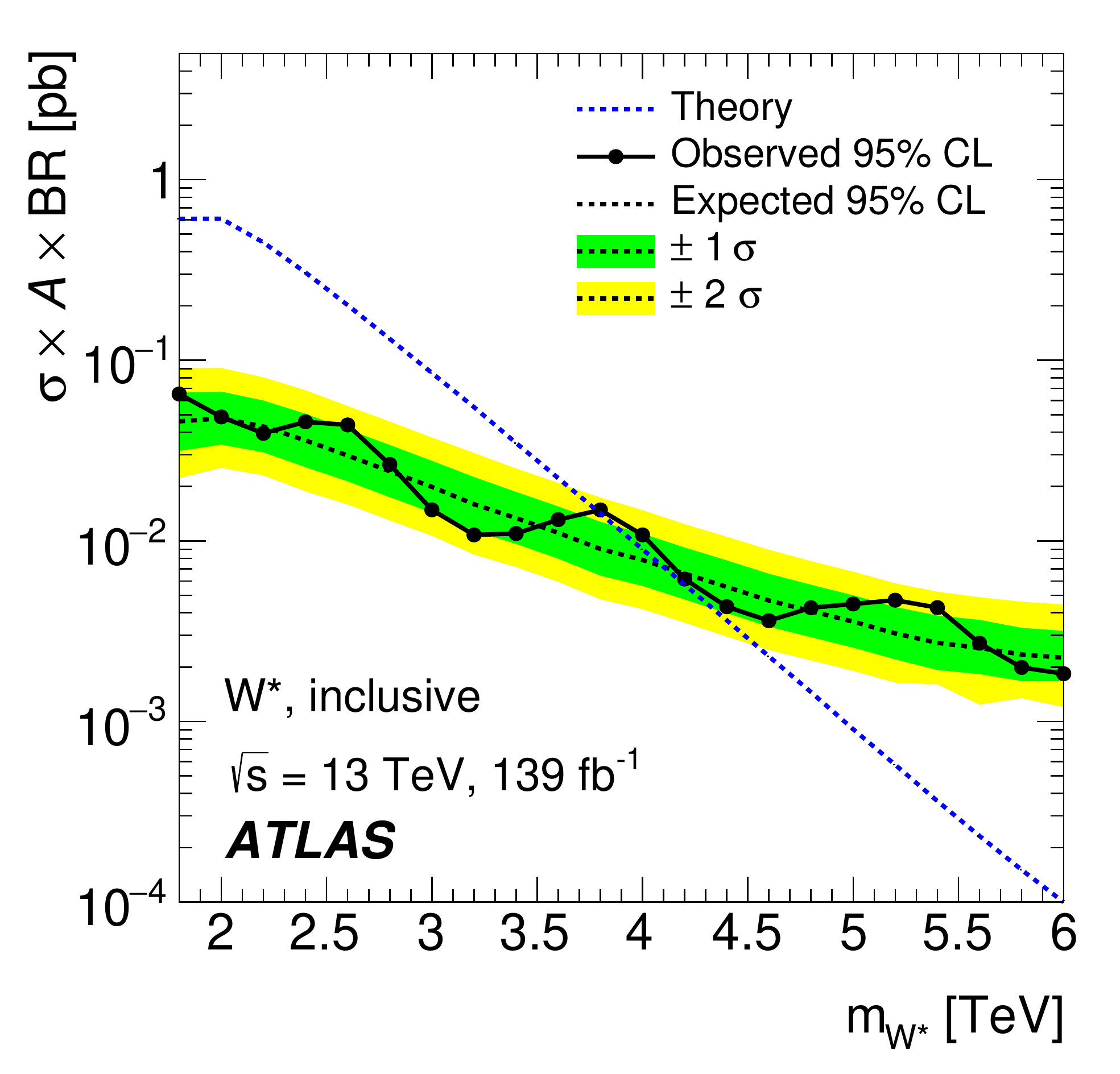}
}
\caption{
The 95\% CL upper limits on the cross-section times acceptance times branching ratio into two jets as a function of the mass of (a) $q^*$, (b) QBH, (c) $W'$ and (d) $W^*$ signals.
The expected upper limit and corresponding $\pm 1\sigma$ and $\pm 2\sigma$ uncertainty bands are also shown.
These exclusion upper limits are obtained using the inclusive dijet selection, with the selection described in the text and summarised in Table~\ref{tab:selection}.
}
\label{fig:inclusive}
\end{figure}
 
\begin{table}[htbp]
\caption{The lower limits on the masses of benchmark signals at 95\% CL.}
\centering
\begin{tabular}{l|c|c|c}
\hline\hline
\multirow{2}{*}{Category} & \multirow{2}{*}{Model} & \multicolumn{2}{c}{Lower limit on signal mass at 95\% CL} \\
& & Observed & Expected \\\hline
\multirow{6}{*}{Inclusive} & $q^*$ & 6.7~\TeV\ & 6.4~\TeV\  \\
& QBH & 9.4~\TeV\  & 9.4~\TeV\ \\
& \Wprime & 4.0~\TeV\ & 4.2~\TeV\  \\
& $W^*$   & 3.9~\TeV\ & 4.1~\TeV\ \\
& DM mediator \Zprime, $g_\text{q}=0.20$ & 3.8~\TeV\ & 3.8~\TeV\ \\
& DM mediator \Zprime, $g_\text{q}=0.50$ & 4.6~\TeV\ & 4.9~\TeV\ \\\hline
$1b$ & $b^*$ & 3.2~\TeV\ & 3.1~\TeV\ \\\hline
\multirow{4}{*}{$2b$} & DM mediator \Zprime $g_\text{q}=0.20$ & 2.8~\TeV\  & 2.8~\TeV\ \\
& DM mediator \Zprime, $g_\text{q}=0.25$ & 2.9~\TeV\  & 3.0~\TeV\ \\
& SSM \Zprime, & 2.7~\TeV\ & 2.7~\TeV\ \\
& graviton, $k/\overline{M}_\text{PL}=0.2$ & 2.8~\TeV\ & 2.9~\TeV\ \\\hline\hline
\end{tabular}
\label{tab:limits}
\end{table}
 
The upper limits obtained from the inclusive category for the signal cross-sections of $q^*$, QBH, \Wprime and $W^*$ are shown in Figure~\ref{fig:inclusive}.
The constraints on the leptophobic DM mediator \Zprime model are shown in Figure~\ref{fig:zprime}.
For the upper limits on the universal coupling $g_\text{q}$ of the \Zprime model, signal points are simulated with 0.5~\TeV\ spacing in mass and spacing as fine as 0.05 in $g_\text{q}$.
A smooth curve is drawn between points by interpolating in $g_\text{q}$ followed by an interpolation in \Zprime mass.
For a given mass, the cross-sections rise with $g_\text{q}$, and thus the upper-left unfilled area is excluded.
The upper limits on the signal yields from the $1b$ category for the $b^*$ signal are shown in Figure~\ref{fig:1b}
and those from the $2b$ category for the $Z'$ and graviton signals are shown in Figure~\ref{fig:2b}.
The lower limits on the signal masses for each of the benchmark models are summarised in Table~\ref{tab:limits}.
For the leptophobic DM mediator \Zprime model the signal constraint from the $2b$ category is comparable to that from the inclusive category at a signal mass of around $1.5~\TeV$, and weaker at higher masses mainly due to the loss of $b$-tagging efficiency. For new states with a larger branching ratio into $b$-quark final states, the $b$-tagged categories will have greater sensitivity.
 
Exclusion upper limits are also set on the cross-section times acceptance times branching fraction into two jets (effective cross-section) of a hypothetical signal modelled as a Gaussian peak in the particle-level \mjj distribution, as shown in Figure~\ref{fig:gauss}.
Gaussian-shaped signal models are tested for different mass hypotheses and various possible signal widths at the detector reconstruction level.
Signal widths range from the detector resolution width of approximately 3\% up to a relative width of $15\%$.
Broader resonances are not considered in this analysis as the presence of the signal would significantly affect the background estimate obtained using the sliding-window fit.
A MC-based transfer matrix connecting the particle-level and reconstruction-level observables is used to fold in the effects of the detector response to the particle-level signals~\cite{EXOT-2016-21}.
For the inclusive category, the upper limits on the effective cross-sections of a Gaussian-shaped signal are approximately 30\text{--}70~fb at a mass of $1.5~\TeV$ and 0.08\text{--}0.2~fb at a mass of $6~\TeV$.
For the $1b$ and $2b$ categories, the upper limits are approximately 5\text{--}20~fb and 4\text{--}6~fb, respectively, at a mass of $1.5~\TeV$.
In the $1b$ category, the highest reach in mass is $5~\TeV$, with upper limits of 0.1\text{--}0.4~fb.
In the $2b$ category, the highest reach in mass is $4.5~\TeV$, with upper limits close to 0.04~fb.
 
\begin{figure}[tbp]
\centering
\includegraphics[width=0.49\textwidth]{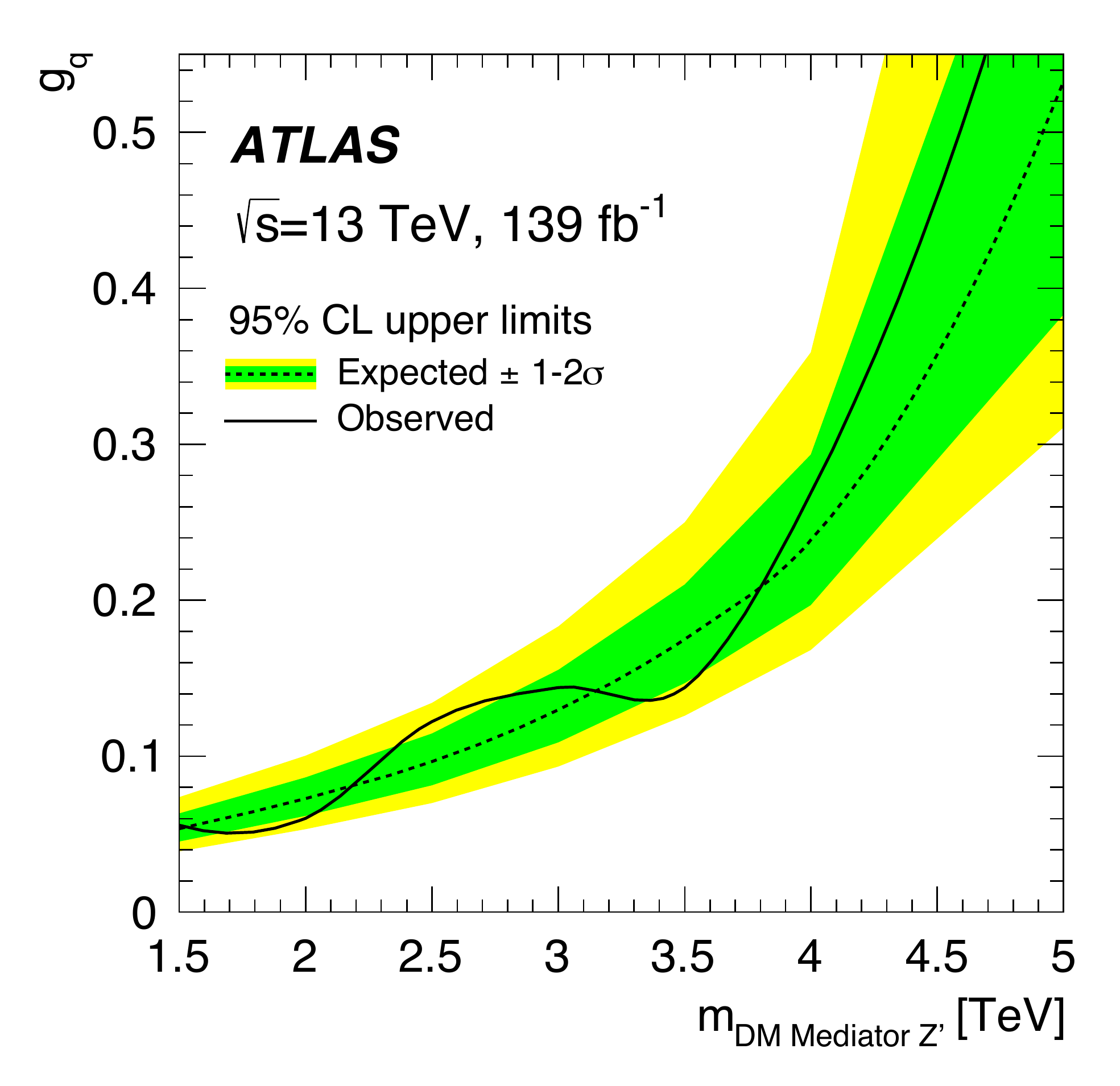}
\caption{
The upper limits on the DM mediator \Zprime signal at 95\% CL from the inclusive category, with the selection described in the text and summarised in Table~\ref{tab:selection}.
The 95\% CL upper limits are set on the universal quark coupling $g_\text{q}$ as a function of the \Zprime mass.
The observed limits (solid) and expected limits (dashed) with $\pm 1\sigma$ and $\pm 2\sigma$ uncertainty bands are shown.
}
\label{fig:zprime}
\end{figure}

\begin{figure}[tbp]
\centering
\includegraphics[width=0.49\textwidth]{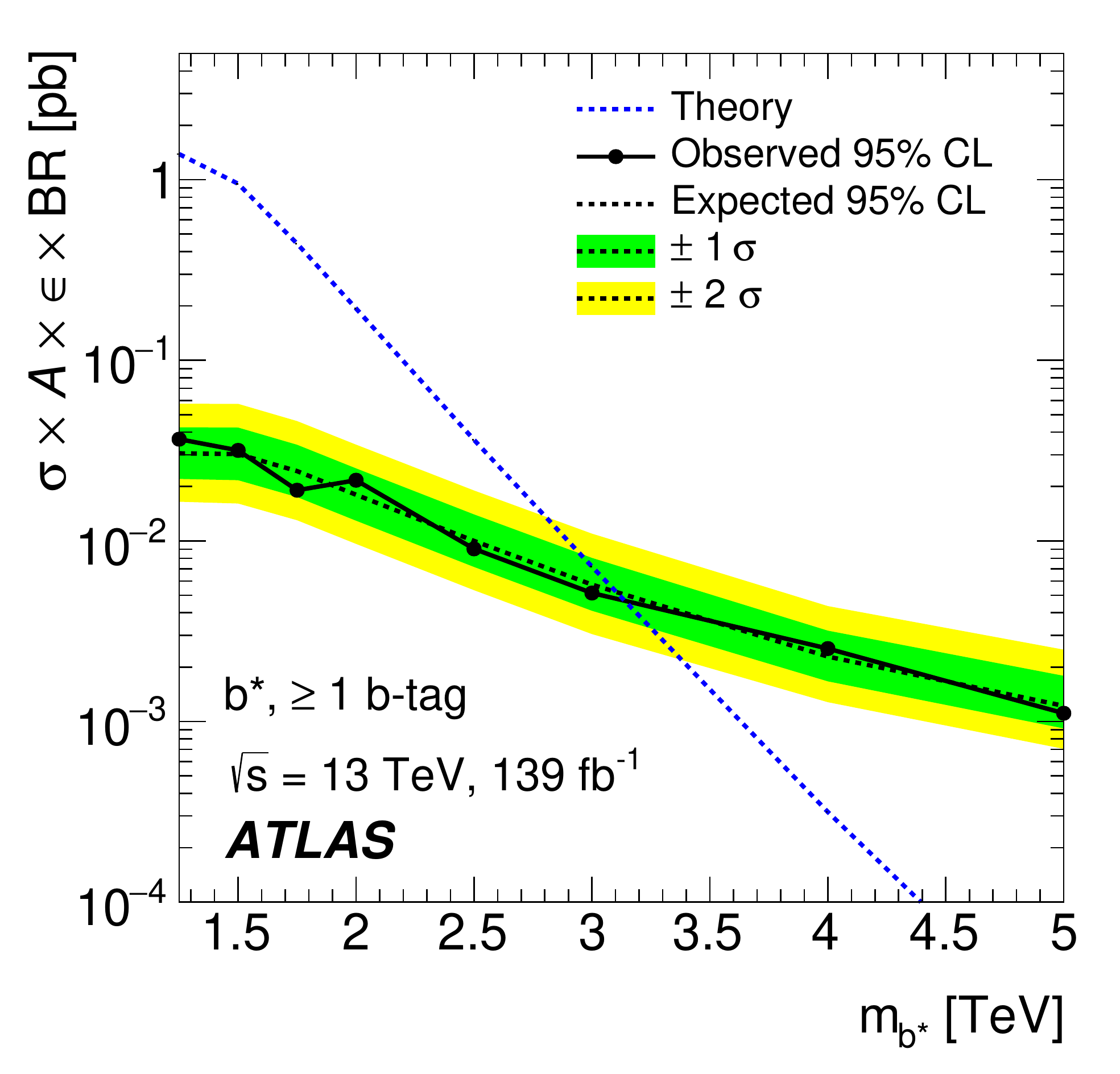}
\caption{
The 95\% CL upper limit on the cross-section times acceptance times $b$-tagging efficiency times branching ratio as a function of the mass of the $b^*$ signal.
The expected limit and corresponding $\pm 1\sigma$ and $\pm 2\sigma$ uncertainty bands are also shown.
These exclusion limits are obtained using the $1b$ category, with the selection described in the text and summarised in Table~\ref{tab:selection}.
}
\label{fig:1b}
\end{figure}
 
\begin{figure}[htbp]
\centering
\subfigure[][]{
\includegraphics[width=0.49\textwidth]{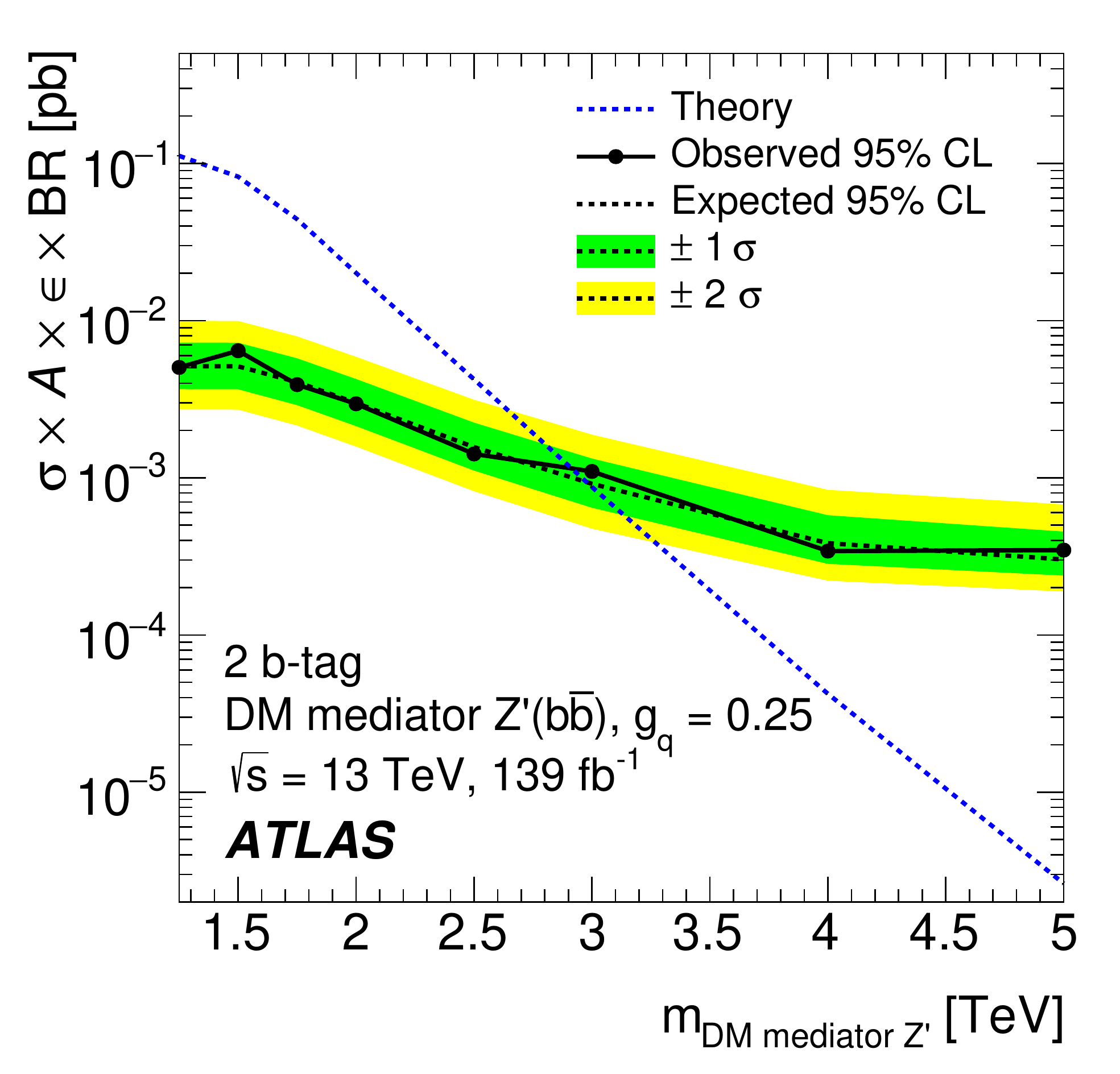}
}\subfigure[][]{
\includegraphics[width=0.49\textwidth]{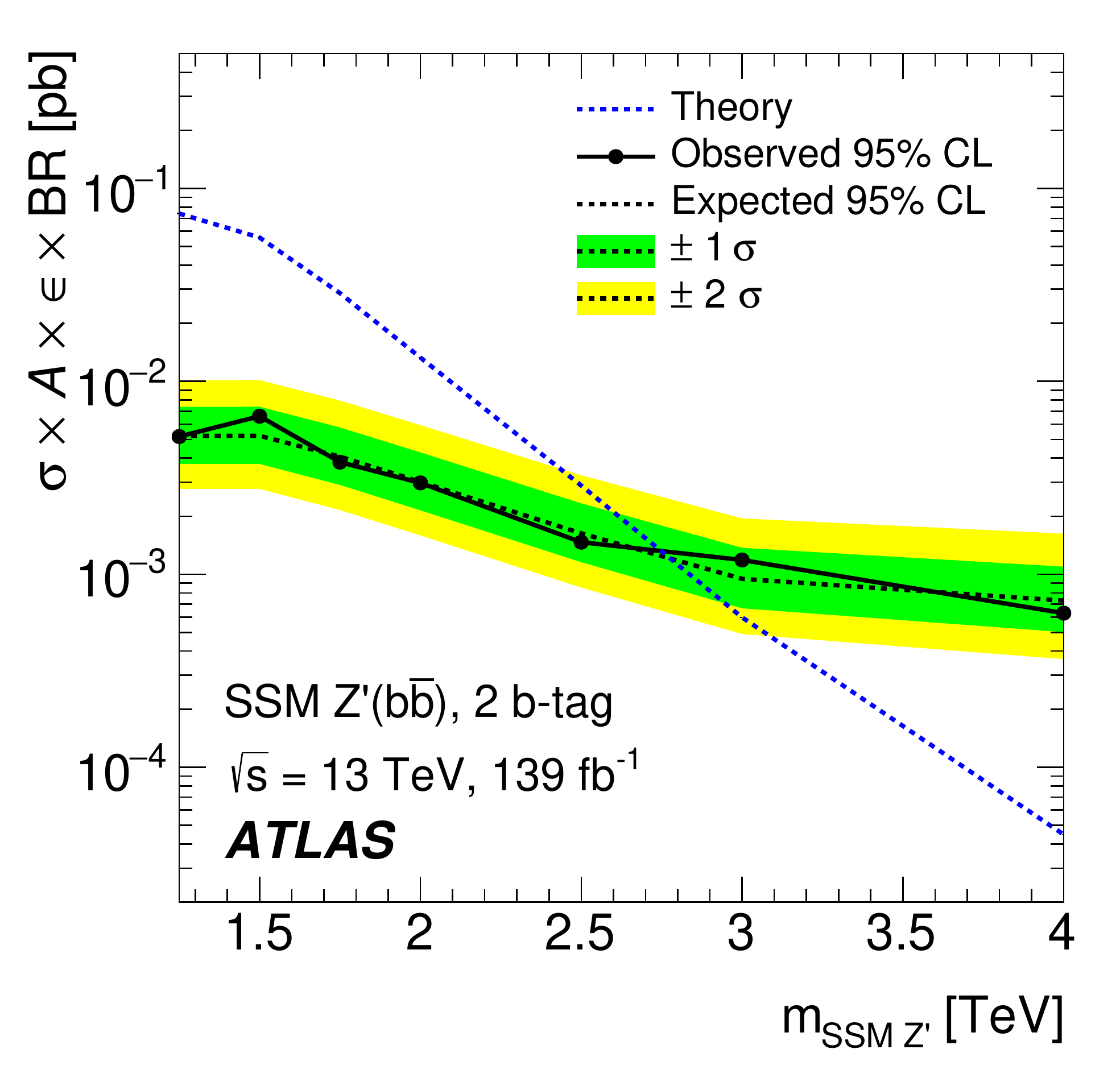}
} \\
\subfigure[][]{
\includegraphics[width=0.49\textwidth]{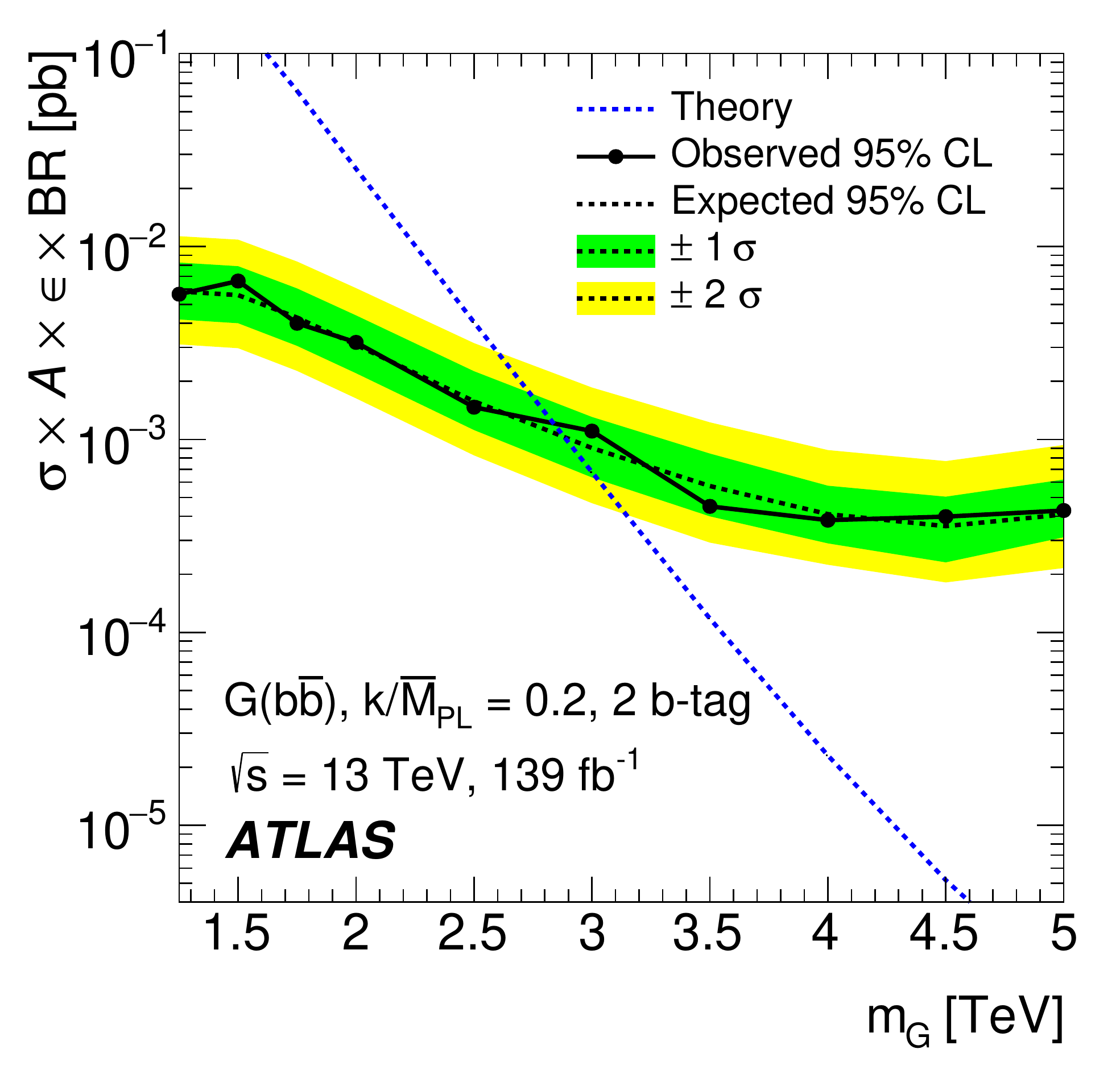}
}
\caption{
The 95\% CL upper limit on the cross-section times acceptance times $b$-tagging efficiency times branching ratio as a function of the signal mass in the (a) DM mediator \Zprime with $g_\text{q}=0.25$, (b) SSM \Zprime and (c) graviton with $k/\overline{M}_\text{PL}=0.2$ models.
The expected limit and corresponding $\pm 1\sigma$ and $\pm 2\sigma$ uncertainty bands are also shown.
These exclusion limits are obtained using the $2b$ category, with the selection described in the text and summarised in Table~\ref{tab:selection}.
}
\label{fig:2b}
\end{figure}
 
\begin{figure}[htbp]
\centering
\subfigure[][]{
\includegraphics[width=0.49\textwidth]{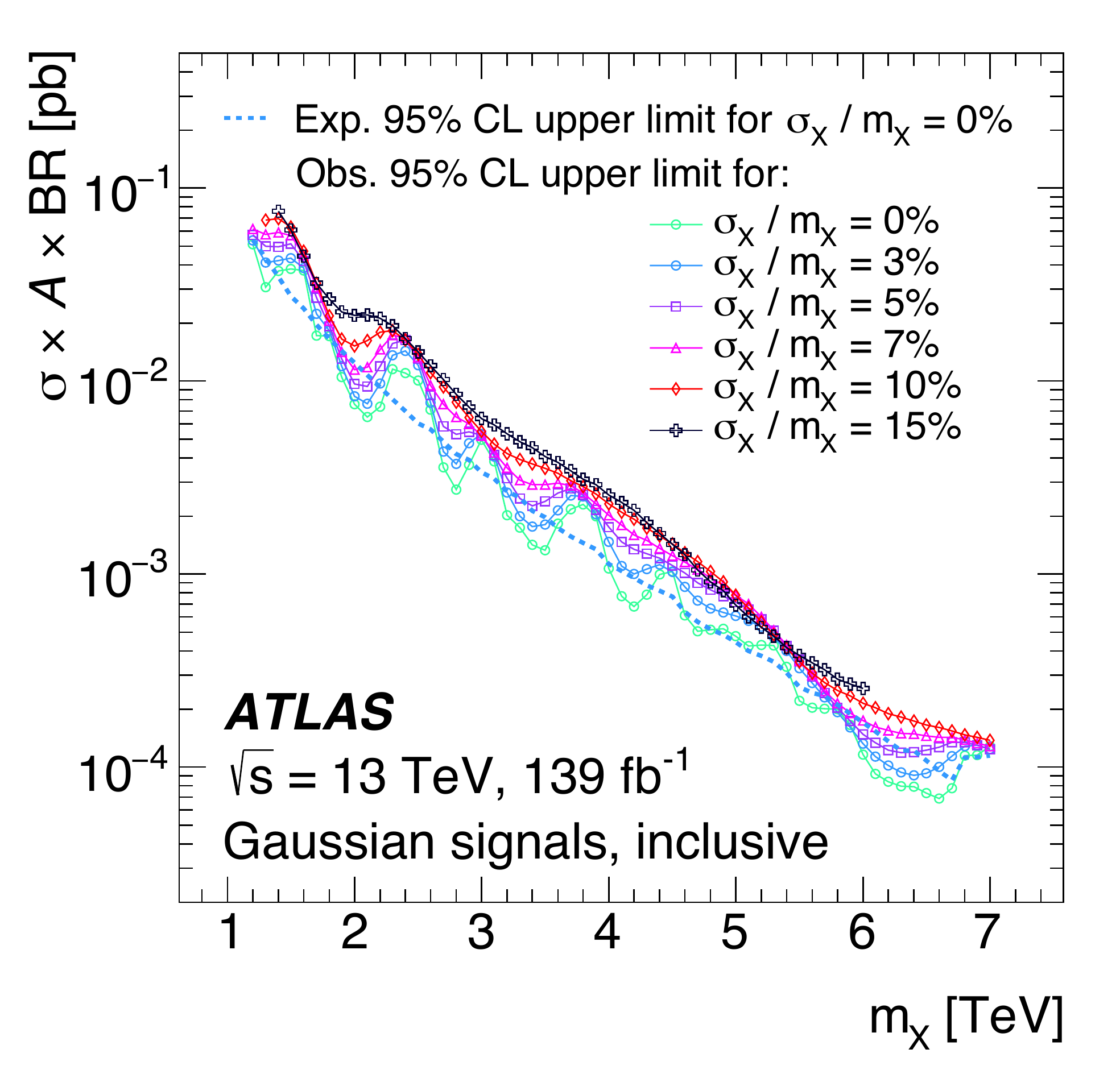}
}\subfigure[][]{
\includegraphics[width=0.49\textwidth]{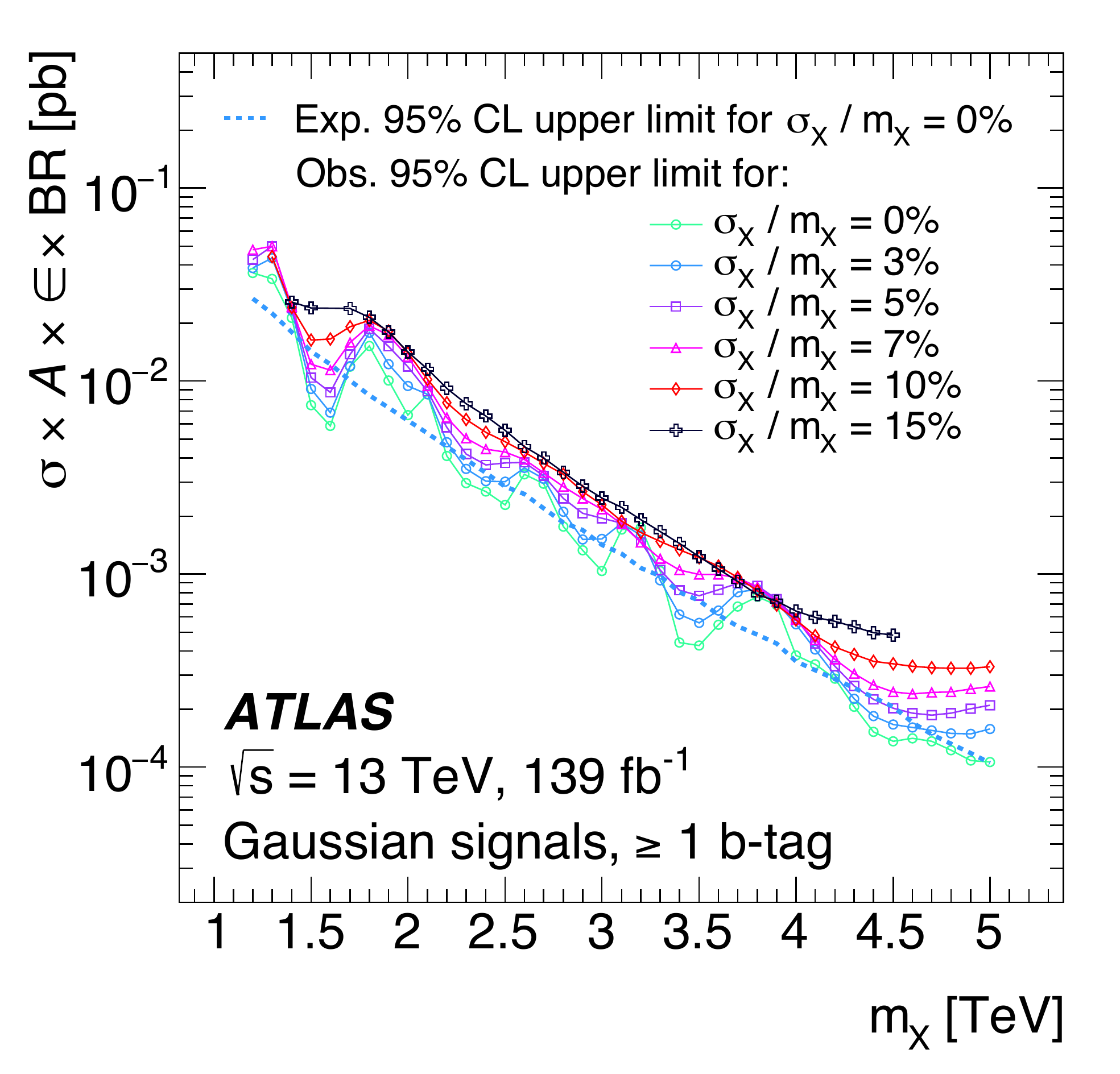}
} \\
\subfigure[][]{
\includegraphics[width=0.49\textwidth]{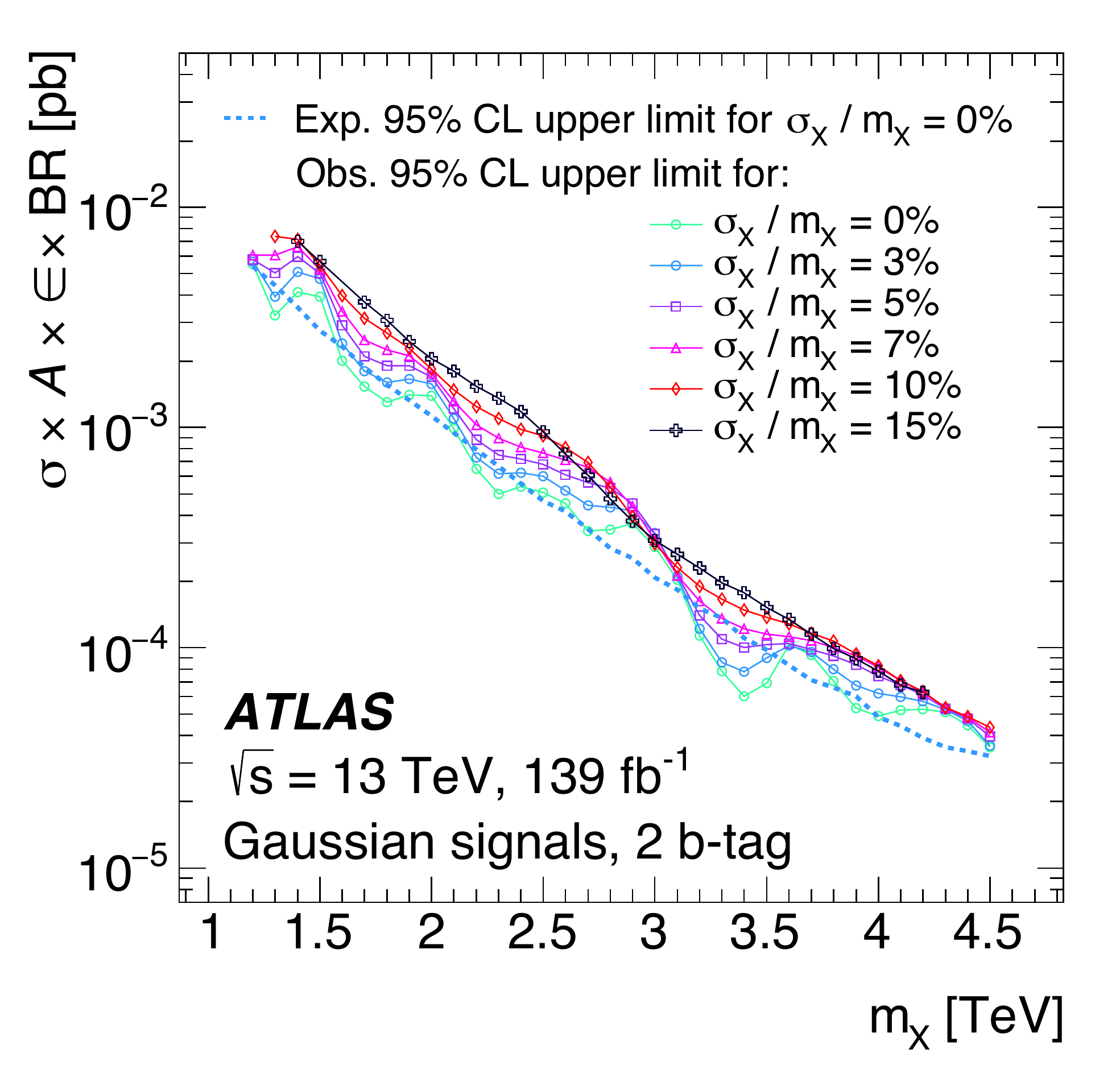}
}
\caption{
The 95\% CL upper limit on the cross-section times kinematic acceptance times branching ratio for resonances with a generic Gaussian shape, as a function of the Gaussian mean mass $m_\text{X}$ in the (a) inclusive, (b) $1b$ and (c) $2b$ categories. For the limits with one or two $b$-jets the $b$-tagging efficiency is included. Different widths, from 0\% up to 15\% of the signal mass, are considered. Gaussian-shape signals with 0\% widths correspond to signal widths smaller than the experimental resolution. For a Gaussian-shaped signal with a relative width of 15\%, the limits are truncated at high mass when the broad signal starts to overlap the upper end of the \mjj spectrum.}
\label{fig:gauss}
\end{figure}
 
The $b$-tagged analysis benefits from substantial improvements in the $b$-jet identification algorithm and associated systematic uncertainties compared with the previous ATLAS result in Ref.~\cite{EXOT-2016-33}.
The current and previous expected 95\% CL upper limits on the cross-section times branching ratio times acceptance times $b$-tagging efficiency are shown in Figure~\ref{fig:money_analysis} as a function of the \Zprime mass in the DM benchmark model. A statistical scaling of the expected upper limits from the previous result (36.1~\ifb) to the current dataset of 139~\ifb\ is also shown, assuming no change to the previous analysis strategy or its uncertainties.
A factor of up to 3.5 improvement beyond that expected from the increase of integrated luminosity in the expected upper limits is observed across the range of masses investigated.
The upper limit of the previous result was obtained with the Bayesian method of Ref.~\cite{Caldwell:2008fw} and with a looser $b$-tagging requirement.
 
\begin{figure}[htbp]
\centering
\includegraphics[width=0.6\textwidth]{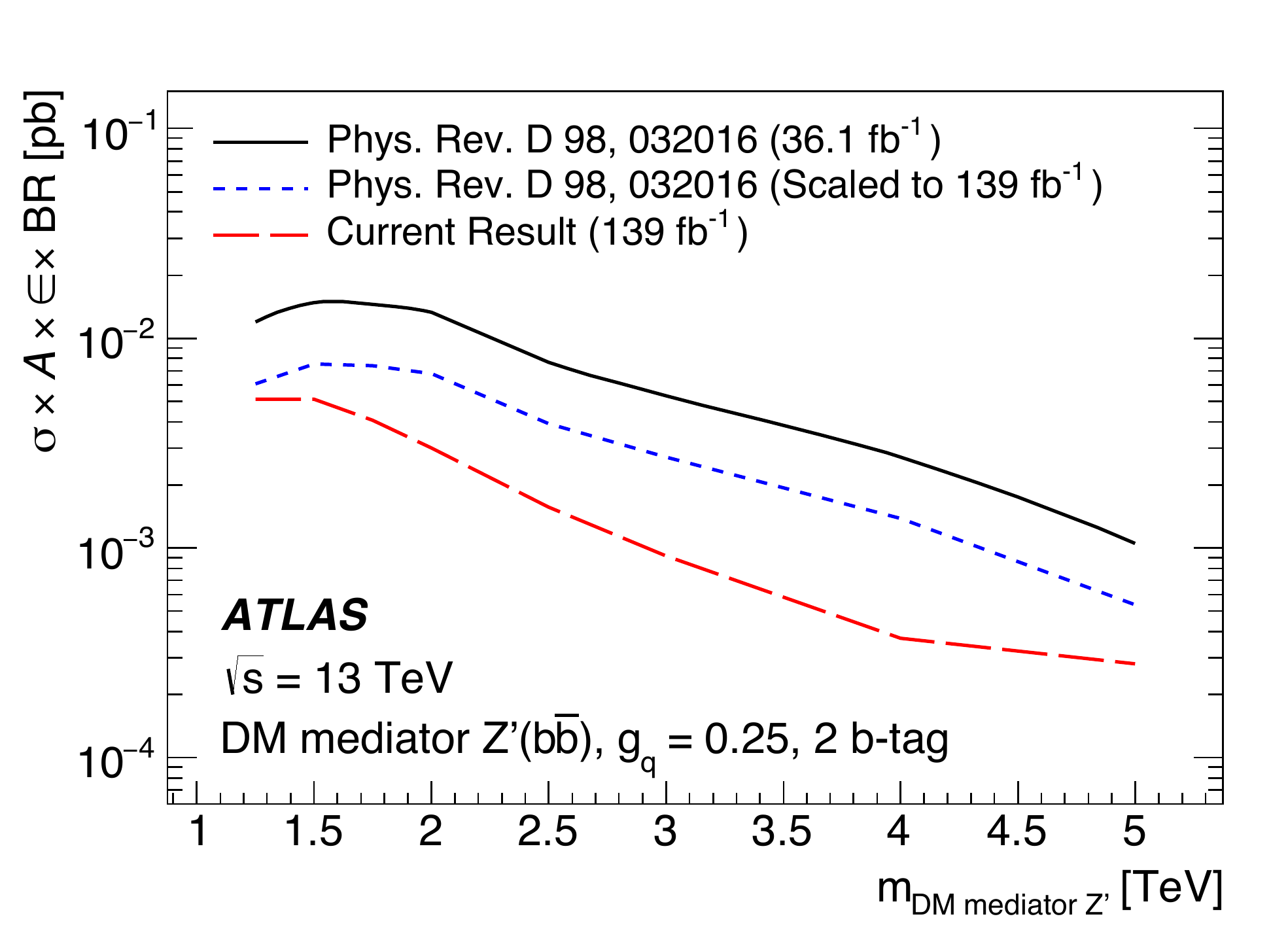}
\caption{
The expected 95\% CL upper limits on the cross-section times acceptance times $b$-tagging efficiency times branching ratio as a function of the DM mediator \Zprime mass for the current and previous iterations of the analysis.
The upper limit of the previous result was obtained with the Bayesian method of Ref.~\cite{Caldwell:2008fw} and is also shown scaled to the 139~\ifb\ integrated luminosity of the current result to illustrate the effect of the analysis improvements.
The current $b$-tagging requirement is tighter than the previous one for high-$\pt$ jets, resulting in a data sample with limited size for \mjj above 4~TeV.
The background rejection, instead, has improved significantly across the entire \mjj spectrum inspected by the analysis.
}
\label{fig:money_analysis}
\end{figure}
 
\FloatBarrier
 
\section{Conclusion}
A search for new resonances decaying into a pair of jets has been performed with dijet events using 139~\ifb\ of proton\text{--}proton collisions recorded at $\sqrt{s}=13~\TeV$ with the ATLAS detector at the Large Hadron Collider between 2015 and 2018.
The invariant mass spectra of the two highest-momentum jets are analysed inclusively, and with at least one or exactly two jets identified as $b$-jets.
No significant excess is observed above the data-driven estimates of the smoothly falling distributions predicted by the Standard Model.
Constraints on various signal models are derived and presented together with model-independent limits on Gaussian-shaped signals.
For example, excited quarks $q^*$ with masses below $6.7~\TeV$ are excluded at 95\% CL. For the SSM \Zprime model, \Zprime masses below $2.7~\TeV$ are excluded at 95\% CL.
The analysis with $b$-tagging benefits from substantial improvements in the $b$-jet identification algorithm at high transverse momentum, resulting in an improvement in sensitivity beyond that expected from the integrated luminosity increase.

\section*{Acknowledgements}
 

We thank CERN for the very successful operation of the LHC, as well as the
support staff from our institutions without whom ATLAS could not be
operated efficiently.
 
We acknowledge the support of ANPCyT, Argentina; YerPhI, Armenia; ARC, Australia; BMWFW and FWF, Austria; ANAS, Azerbaijan; SSTC, Belarus; CNPq and FAPESP, Brazil; NSERC, NRC and CFI, Canada; CERN; CONICYT, Chile; CAS, MOST and NSFC, China; COLCIENCIAS, Colombia; MSMT CR, MPO CR and VSC CR, Czech Republic; DNRF and DNSRC, Denmark; IN2P3-CNRS and CEA-DRF/IRFU, France; SRNSFG, Georgia; BMBF, HGF and MPG, Germany; GSRT, Greece; RGC and Hong Kong SAR, China; ISF and Benoziyo Center, Israel; INFN, Italy; MEXT and JSPS, Japan; CNRST, Morocco; NWO, Netherlands; RCN, Norway; MNiSW and NCN, Poland; FCT, Portugal; MNE/IFA, Romania; MES of Russia and NRC KI, Russia Federation; JINR; MESTD, Serbia; MSSR, Slovakia; ARRS and MIZ\v{S}, Slovenia; DST/NRF, South Africa; MINECO, Spain; SRC and Wallenberg Foundation, Sweden; SERI, SNSF and Cantons of Bern and Geneva, Switzerland; MOST, Taiwan; TAEK, Turkey; STFC, United Kingdom; DOE and NSF, United States of America. In addition, individual groups and members have received support from BCKDF, CANARIE, Compute Canada and CRC, Canada; ERC, ERDF, Horizon 2020, Marie Sk{\l}odowska-Curie Actions and COST, European Union; Investissements d'Avenir Labex, Investissements d'Avenir Idex and ANR, France; DFG and AvH Foundation, Germany; Herakleitos, Thales and Aristeia programmes co-financed by EU-ESF and the Greek NSRF, Greece; BSF-NSF and GIF, Israel; CERCA Programme Generalitat de Catalunya and PROMETEO Programme Generalitat Valenciana, Spain; G\"{o}ran Gustafssons Stiftelse, Sweden; The Royal Society and Leverhulme Trust, United Kingdom.
 
The crucial computing support from all WLCG partners is acknowledged gratefully, in particular from CERN, the ATLAS Tier-1 facilities at TRIUMF (Canada), NDGF (Denmark, Norway, Sweden), CC-IN2P3 (France), KIT/GridKA (Germany), INFN-CNAF (Italy), NL-T1 (Netherlands), PIC (Spain), ASGC (Taiwan), RAL (UK) and BNL (USA), the Tier-2 facilities worldwide and large non-WLCG resource providers. Major contributors of computing resources are listed in Ref.~\cite{ATL-SOFT-PUB-2020-001}.
 

\clearpage

\printbibliography

\clearpage \input{atlas_authlist}

\end{document}

%% file: atlas_authlist.tex
 
\begin{flushleft}
\hypersetup{urlcolor=black}
{\Large The ATLAS Collaboration}

\bigskip

\AtlasOrcid[0000-0002-6665-4934]{G.~Aad}$^\textrm{\scriptsize 102}$,    
\AtlasOrcid[0000-0002-5888-2734]{B.~Abbott}$^\textrm{\scriptsize 129}$,    
\AtlasOrcid{D.C.~Abbott}$^\textrm{\scriptsize 103}$,    
\AtlasOrcid[0000-0002-2788-3822]{A.~Abed~Abud}$^\textrm{\scriptsize 71a,71b}$,    
\AtlasOrcid[0000-0002-1002-1652]{K.~Abeling}$^\textrm{\scriptsize 53}$,    
\AtlasOrcid[0000-0002-2987-4006]{D.K.~Abhayasinghe}$^\textrm{\scriptsize 94}$,    
\AtlasOrcid[0000-0002-8496-9294]{S.H.~Abidi}$^\textrm{\scriptsize 167}$,    
\AtlasOrcid[0000-0002-8279-9324]{O.S.~AbouZeid}$^\textrm{\scriptsize 40}$,    
\AtlasOrcid{N.L.~Abraham}$^\textrm{\scriptsize 156}$,    
\AtlasOrcid[0000-0001-5329-6640]{H.~Abramowicz}$^\textrm{\scriptsize 161}$
% ...
\bigskip
\\

% ...
$^{40}$Niels Bohr Institute, University of Copenhagen, Copenhagen; Denmark.\\
% ...
$^{53}$II. Physikalisches Institut, Georg-August-Universit\"{a}t G\"ottingen, G\"ottingen; Germany.\\
% ...
$^{71}$$^{(a)}$INFN Sezione di Pavia;$^{(b)}$Dipartimento di Fisica, Universit\`a di Pavia, Pavia; Italy.\\
% ...
$^{94}$Department of Physics, Royal Holloway University of London, Egham; United Kingdom.\\
% ...
$^{102}$CPPM, Aix-Marseille Universit\'e, CNRS/IN2P3, Marseille; France.\\
$^{103}$Department of Physics, University of Massachusetts, Amherst MA; United States of America.\\
% ...
$^{129}$Homer L. Dodge Department of Physics and Astronomy, University of Oklahoma, Norman OK; United States of America.\\
% ...
$^{156}$Department of Physics and Astronomy, University of Sussex, Brighton; United Kingdom.\\
% ...
$^{161}$Raymond and Beverly Sackler School of Physics and Astronomy, Tel Aviv University, Tel Aviv; Israel.\\
% ...
$^{167}$Department of Physics, University of Toronto, Toronto ON; Canada.\\
% ...
$^{a}$ Also at Borough of Manhattan Community College, City University of New York, New York NY; United States of America.\\
$^{b}$ Also at CERN, Geneva; Switzerland.\\
% ...

\end{flushleft}
